\documentclass[a4paper,fleqn,usenatbib]{mnras}

\usepackage{graphicx}
\usepackage{amssymb}
\usepackage{amsmath}
\usepackage{url}
\usepackage{bm}
\usepackage{aas_macros}
\usepackage{rotating}
\usepackage{booktabs}
\usepackage{braket}
\usepackage[neverdecrease]{paralist}
\usepackage{xurl}
\usepackage{hhline}
\usepackage{array}    
\usepackage{ragged2e} 
\usepackage{tabularx}
\usepackage{booktabs}

\newcommand{\ks}{K_{\text s}} 
\usepackage{tabularray}
\usepackage{multirow}
\usepackage{color} 

\title[]{An automated method to detect and characterise semi-resolved star clusters}

\author[A.~E.~Miller et al.]{Amy E.~Miller,${^{1,2,3,4}}$\thanks{E-mail:
  amyelizmiller@gmail.com (AEM)} Zachary Slepian,${^{5}}$  Elizabeth A. Lada,${^{5}}$  Richard de Grijs,$^{1,2,6}$
  \newauthor
  Maria-Rosa L. Cioni,$^{3}$ Mark R. Krumholz,$^{7,8}$ Amir E. Bazkiaei,$^{2,8,9}$ 
  \newauthor
  Valentin D. Ivanov,$^{10}$ Joana M. Oliveira,$^{11}$ Vincenzo Ripepi,$^{12}$ and 
  \newauthor
  Jacco Th. van Loon$^{11}$\\
 \\
  $^1$ School of Mathematical and Physical Sciences, Macquarie University, Balaclava Road, Sydney, NSW 2109, Australia\\
  $^2$ Astrophysics and Space Technologies Research Centre, Macquarie University, Balaclava Road, Sydney, NSW 2109, Australia\\
  $^3$ Leibniz-Institut f{\"u}r Astrophysik Potsdam, An der Sternwarte 16, D-14482 Potsdam, Germany\\
  $^4$ Institut f\"{u}r Physik und Astronomie, Universit\"{a}t Potsdam, Haus 28, Karl-Liebknecht-Str. 24/25, D-14476 Potsdam, Germany\\
  $^5$ Department of Astronomy, University of Florida, Gainesville, FL 32611, USA\\
  $^6$ International Space Science Institute--Beijing, 1 Nanertiao, Zhongguancun, Hai Dian District, Beijing 100190, China\\
  $^7$ Research School of Astronomy and Astrophysics, Australian National University, Canberra, ACT 2611, Australia\\
  $^8$ Australian Research Council Centre of Excellence in All-Sky Astrophysics (ASTRO3D), Canberra, ACT 2611, Australia\\
  $^9$ Australian Astronomical Optics, Macquarie University, North Ryde, NSW 2113, Australia \\
  $^{10}$European Southern Observatory, Karl-Schwarzschild-Str. 2, D-85748 Garching bei M\"{u}nchen, Germany\\
  $^{11}$Lennard-Jones Laboratories, Keele University, ST5 5BG, UK\\
  $^{12}$INAF -- Osservatorio Astronomico di Capodimonte, Salita Moiariello 16, I--80131 Naples, Italy\\
  }

\begin{document}

\date{Accepted ?, Received ?; in original form ?}

\pagerange{\pageref{firstpage}--\pageref{lastpage}} \pubyear{2024}

\maketitle

\label{firstpage}

\begin{abstract}
We present a novel method for automatically detecting and characterising semi-resolved star clusters: clusters where the observational point-spread function (PSF) is smaller than the cluster's radius, but larger than the separations between individual stars. We apply our method to a 1.77 deg$^2$ field located in the Large Magellanic Cloud (LMC) using the VISTA survey of the Magellanic Clouds (VMC), which surveyed the LMC in the $YJK_\text{s}$ bands. Our approach first models the position-dependent PSF to detect and remove point sources from deep $K_\text{s}$ images; this leaves behind extended objects such as star clusters and background galaxies. We then analyse the isophotes of these extended objects to characterise their properties, perform integrated photometry, and finally remove any spurious objects this procedure identifies. We demonstrate our approach in practice on a deep VMC $\ks$ tile that contains the most active star-forming regions in the LMC: 30 Doradus, N158, N159, and N160. We select this tile because it is the most challenging for automated techniques due both to crowding and nebular emission. We detect 682 candidate star clusters, with an estimated contamination rate of 13\% from background galaxies and chance blends of physically unrelated stars. We compare our candidates to publicly available {\sl James Webb Space Telescope} data and find that at least 80\% of our detections appear to be star clusters. 

\end{abstract}


\section{Introduction}
\label{introduction}
Most stars form within hierarchical groupings \citep{lada1991b,lada1992,lada&lada2003,sun2018smc,krumholz2019,miller2022} that include (in ascending order in size and descending order in density) clusters, associations, and complexes. Young star clusters are the building blocks of this hierarchy. However, despite their central role, we still have limited understanding of the processes involved in their formation and of their physical properties, as well as how these factors depend on galactic environment. This limited understanding in part stems from the fact that detailed studies of such clusters have only been conducted in the Milky Way \citep[\textit{e.g.}][]{lada1995,Muench2002,gutermuth2009}. This means that the dependence of the initial mass function \citep[\textit{e.g.}][]{Schneider2018,Gennaro2020} and star-forming scaling relations, such as the Kennicutt--Schmidt relation \citep{lada2010,lada2012,krumholz2015}, on galactic environment is not well understood. 

The Large Magellanic Cloud (LMC) is a particularly appealing target for expanding our young cluster sample to a different galaxy. It is nearby \citep[50 kpc;][]{degrijs2014}, and its face-on orientation and small line-of-sight depth \citep{subramanian2009} remove the line-of-sight confusion and distance uncertainty that hamper studies of star clusters in the Milky Way. The LMC features vigorous star formation, including one of the most active star-forming regions in the local Universe, the 30 Doradus complex \citep{Fahrion2023}. The LMC is also interacting with its companion, the Small Magellanic Cloud (SMC); thus the LMC is a unique laboratory for the study of interaction-driven star formation. The LMC's sub-solar metallicity \citep[$Z \sim 0.5  Z_\odot$;][]{rolleston2002} also makes it an interesting target for the study of star formation in the metal-poor environment that prevailed in the early Universe, as the interstellar medium during the epoch of peak star formation around $z\sim1.5$ is similar to that of the LMC \citep{Pei1999}. 

While the LMC is an attractive target, it presents challenges as well as opportunities. Using ground-based telescopes, many young star clusters in the LMC are semi-resolved. Before going on, we present several crucial observational definitions of star clusters. We define the semi-resolved cluster regime loosely as one where angular separations between individual stars in the cluster are much smaller than the observational point-spread function (PSF), but the cluster angular radius is at least twice as large than the PSF. In contrast, we define the barely-resolved regime as one where the cluster radius is comparable to the PSF, and the unresolved regime as one where clusters are indistinguishable from point sources because their radii are smaller than the PSF. Star clusters in the Milky Way, the LMC, the SMC, M31, and M33 can all lie in the semi-resolved regime depending on the telescope used and the stellar density and luminosity distribution of the region being studied.

To understand why the semi-resolved regime is challenging, we will now briefly outline how star clusters are identified in imaging data. The most common approach to this problem in the Milky Way is to measure individual stars' positions and then apply cluster-finding methods or similar spatial statistics to the resulting collection of point positions \citep[\textit{e.g.}][]{carpenter2000, ivanov2002}. Several works have used this approach in the LMC, for instance, \citet{sun201730dor,sun2017bar,sun2018smc}, \citet{zivkov2018} and \citet{miller2022} all studied clustering using the positions of individual pre- or upper main-sequence stars. However, crowding causes spatial-statistics-based approaches to fail in compact, clustered regions. 

We can see that many LMC regions are affected by crowding as follows. Stars in young Milky Way clusters are separated by $< 0.1$ pc, while the PSF size for large, ground-based optical and near-infrared photometric surveys is typically limited by seeing to $\sim$1 arcsec, corresponding to $\approx 0.24$ pc at the LMC's distance. Assuming that interstellar separations in young clusters of the LMC are similar to those in the Milky Way, this means that we cannot resolve individual LMC stars. Thus, any method that requires individual stars' positions (as spatial clustering statistics do) is bound to break down. 

At the same time, the automated cluster recovery techniques employed to find star clusters in more distant galaxies by surveys such as the Legacy Extragalactic Ultraviolet Survey \citep[LEGUS;][]{krumholz2015,messa2018,cook2019} and the Physics At High Angular resolution in Nearby Galaxies survey \citep[PHANGS;][]{whitmore2021,thilker2022} are also not suitable for the LMC. At the typical distances of LEGUS and PHANGS targets ($\sim3$--10 Mpc), star clusters are barely-resolved. Consequently, star clusters can be identified by detecting bright near-point sources that are slightly more extended than the observational PSF. However, this method would not work for ground-based observations of the LMC because the clusters we can resolve and detect are generally much more extended than the PSF. The improved angular resolution and sensitivity of our observations allow us to identify more complex structures.

In addition to the challenges posed by lack of resolution, the nebular structures associated with star formation---molecular and ionised gas clouds (H$_{2}$ and H\,{\sc ii} regions, respectively)---present a further complication. Wide-scale surveys of stars are usually conducted at near-infrared, optical, and ultraviolet wavelengths; recent examples in the LMC include the VISTA Survey of the Magellanic Clouds \citep[VMC;][]{cioni2011} in $YJ\ks$, the Survey of the Magellanic Stellar History \citep[SMASH;][]{nidever2021} in $ugriz$, and the Galaxy Evolution Explorer (GALEX) survey of the Magellanic system \citep{Simons2014} in the near- and far-ultraviolet. In young star-forming regions, nebular structures block ultraviolet and optical light, scatter ultraviolet, optical and near-infrared light, produce line emission at optical, ultraviolet and near-infrared wavelengths, and absorb and re-emit light in the near-infrared \citep{ferriere2001,draine2003}. Therefore, young clusters identified via these passbands are almost always accompanied by nebular emission and variable extinction which both make detecting and characterising them even more difficult. 

In short, the challenges of identifying young clusters in nearby neighboring galaxies are several, including limited angular resolution, crowding, and confusion due to nebular emission.

Consequently, star clusters in the Magellanic Clouds, M31 and M33, and even in more distant and obscured parts of the Milky Way, have often been found by eye. For example, \citet{dutra2000} searched $\ks$ Milky Way images images near the Galactic Centre for extended objects with angular sizes of at least 1 arcmin. \citet{bica1995,1999bica,bica2008} analyzed previous LMC star cluster catalogues and visually identified the clusters as extended objects in optical photographic plates. Additionally, they identified new cluster candidates by eye. \citet{romita2016} also adopted a by-eye approach to finding embedded star clusters in deep $K_\text{s}$ images of the LMC, focusing on over-densities that overlap with CO emission, and thus are presumably young, having not yet completely cleared their natal molecular clouds. Finally, the Panchromatic Hubble Andromeda Treasury (PHAT) and the Panchromatic Hubble Andromeda Treasury: Triangulum Extended Region (PHATTER) surveys of M31 and M33, which lie in the same distance/resolution regime as ground-based LMC surveys, adopted a crowd-sourced approach by having thousands of volunteers identify and classify star clusters by eye \citep{Johnson2015,Johnson2022}.  


Cluster finding by eye can work reasonably well, especially when seeking obvious, massive cluster candidates in imaging data. However, by-eye identification and classification can be difficult to scale and replicate. Even worse, the by-eye approach can introduce unforeseen biases in the detected sample. 

Large, crowd-sourced searches have been conducted to find and classify objects such as star clusters and galaxies. One such platform, Galaxy Zoo, had volunteers classify galaxy morphologies using Sloan Digital Sky Survey images \citep{Lintott2008}. Initially, the findings indicated that the Universe has a preference for right-handed spiral galaxies over left-handed ones. However, this turned out to be a bias in human perception rather than in nature \citep{Land2008}. The results from Galaxy Zoo illustrate that volunteer by-eye classification can introduce subtle biases that can then propagate into, for example, training sets for machine-learning algorithms. 


Therefore, our paper presents a novel, fully automated method for detecting and characterizing semi-resolved star clusters based on their extended shapes compared to the observational PSF. Leveraging this difference in shape, we model the PSF of near-infrared images and use our model to detect and remove point-sources. The leftover objects in the image are extended objects such as star clusters, background galaxies, or similar extended objects. We then detect objects and characterise them using an isophotal analysis, by analyzing the isophotes at different significance levels.

\begin{figure*}
  \includegraphics[width=.8\linewidth]{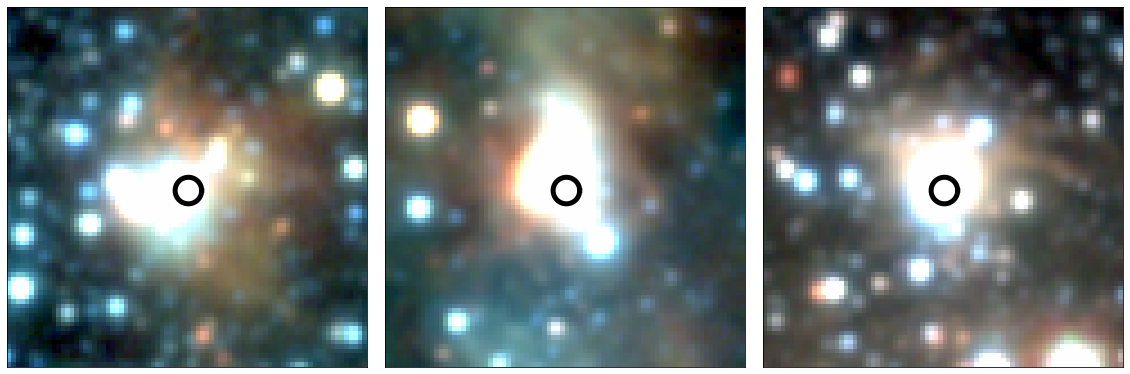}
  \caption{Deep $YJK_\text{s}$ RGB images of sample cluster candidates from \citet{romita2016}. The black circle indicates the PSF FWHM of 1 arcsec. The images illustrate the semi-resolved star clusters that our method is designed to find and characterise, as described in \S\ref{introduction}. We define the semi-resolved star cluster regime as when the angular separation between stars is much less than the PSF size, but where the cluster radii are roughly two or more times larger than the PSF.}
  \label{fig:kr_clusters}
\end{figure*}

To remove stars from near-infrared images and detect semi-resolved star clusters, we use data obtained from the VISTA Survey of the Magellanic Clouds \citep[VMC;][]{cioni2011}. The VMC data cover the main bodies of the LMC and SMC, as well as regions in the Magellanic Bridge and Stream, in $YJ \ks$. The VMC is characterised by seeing $\sim$1 arcsec; therefore, small young star clusters in the LMC should be semi-resolved. We show three examples of the type of semi-resolved star clusters of interest for this study in Figure \ref{fig:kr_clusters}. Our method is designed to automatically detect and characterise sources that look like these.

In this paper we perform a pilot study of one 1.77 deg$^2$ VMC region in the LMC where the star-forming regions 30 Doradus, N158, N159 and N160 are located. This region is a superb test case because it is the most nebulous \citep[hosting many molecular clouds and H\,{\sc ii} regions;][]{kennicutt&hodge1986,wong2011} and busiest region \citep{sun201730dor, Fahrion2023} covered by VMC data. Therefore, showing that our method works in this region instills confidence that it will work on every other region the VMC data cover. 

This paper is structured as follows: In \S\ref{sec:VMC}, we introduce the VMC data. Following this, in \S\ref{sec:data_reduction}, we outline our pipeline for data reduction. Subsequently, we model the PSF in \S\ref{sec:modeling_psf}, and in \S\ref{sec:creation_residual_image}, we utilize our model to remove point sources and generate the residual image. Moving forward, in \S\ref{sec:isophotes}, we employ an isophotal analysis to detect extended objects on the residual image, followed by a culling procedure and presentation of the candidate clusters. Lastly, in \S\ref{sec:discussion}, we compare our findings with previous studies and publicly available {\sl James Webb Space Telescope} ({\sl JWST}) images. We conclude our work in \S\ref{sec:conclusion}.

\section{VMC Data}
\label{sec:VMC}

The VMC survey \citep{cioni2011} mapped the Magellanic System in the near-infrared $YJ \ks$ bands over 170 deg$^2$. The data were taken between 2009 and 2018 using the VISTA infrared camera \citep[VIRCAM;][]{dalton2006} on the 4.1 m VISTA telescope \citep{sutherland2015}, operated by the European Southern Observatory (ESO). VIRCAM has 16 Raytheon VIRGO detectors arranged in a $4 \times 4$ array. Each detector is $2048 \times 2048$ pixels in size and is separated by 42.5\% in $X$ and 90\% in $Y$ from their nearest neighbours. The mean pixel scale is 0.339 arcsec, which corresponds to $\sim 0.08$ pc at the distance of the LMC. Five jitter (also known as dither) exposures are captured at each pointing to remove cosmetic defects. These are image artefacts that are intrinsic to the cosmetic properties of the VIRCAM detectors. Each exposure is made of 4, 8, or 15 Detector Integration Time images (DITs), depending on waveband, where the integration time per DIT corresponds to 20, 10, and 5 s in the $Y$, $J$, and $K_\text{s}$ bands, respectively. The stacked jitter exposures are called `pawprints.'

Six pawprints are taken in succession in a specific pointing pattern to produce a contiguous image, which is called a `tile.' Each tile covers $\sim$1.77 deg$^2$, including two underexposed wing regions where the exposure time is about half that of the inner $\sim$1.5 deg$^2$. Pawprints obtained at multiple epochs are further stacked to create `deep' pawprints. In the VMC survey there are at least four epochs in the $Y$ and $J$ bands and 13 in $K_\text{s}$; two epochs per band are shallow. Deep pawprints are mosaicked to create deep tiles that homogeneously cover the spatial holes among the pawprints. There are 68 tiles that cover the LMC, the locations of which are labelled by two numbers indicating the relevant row and column within the area of the galaxy. We refer the reader to \cite{cioni2011} for further details on the VMC observing strategy.

The Cambridge Astronomical Survey Unit (CASU) delivers reduced pawprints, tiles, and catalogues. The Wide Field Astronomy Unit (WFAU) at the University of Edinburgh receives the calibrated images and catalogues and ingests them into the VISTA Science Archive\footnote{\url{http://horus.roe.ac.uk/vsa/}} \citep[VSA;][]{cross2012}. The VSA then generates deeper images and catalogues from combined multi-epoch images. The data are reduced using the VISTA Data Flow System \citep[VDFS;][]{irwin2004,gonzalez2017}, version 1.5.\footnote{\url{http://casu.ast.cam.ac.uk/surveys-projects/vista/vdfs}}

As we present an image reduction pipeline in the following sections, we briefly describe the image data reduction steps:
\begin{enumerate}[\bfseries (i)]
    \item The VDFS subtracts dark current from the raw image and corrects linearity.
    \item The VDFS applies a flat-field correction by dividing by twilight flat fields.
    \item The VDFS models and subtracts a sky background from each exposure. Various algorithms are employed to estimate the 2D background model, which involve integrating the science images with pixel rejection or masking techniques. VDFS subtracts the sky background, and then adds the mean back to the image.
    \item In order to make pawprints, CASU performs jitter stacking via bilinear interpolation after they correct for the stripe pattern produced by the detector readout electronics.
    \item Next, CASU mosaics the pawprints into tiles using another bilinear interpolation routine in a process that corrects for astrometric and photometric distortions.
    \item WFAU performs the stacking and mosaicking of the deep, multi-epoch images in the same manner.
    \item At each pawprint or tile creation stage, CASU or WFAU generate a catalogue after applying a nebulosity filter that removes large-scale variations (at the level of $>$30 arcsec) in the image. The images with the nebulosity filter applied are, however, not kept because this is only a step to optimise the detection of point sources. 
    \item CASU and WFAU use the catalogues to recalculate the zero points, extinction and other header keywords pertaining to the deeper images.
\end{enumerate}
More information about the data reduction and software used is available from the CASU webpages.\footnote{\url{http://casu.ast.cam.ac.uk/surveys-projects/vista}}

The VMC team created the PSF catalogues from the pawprints. The procedure is to find a mean PSF model for every detector image and for every available epoch. Then, homogenisation of all detector images to the largest PSF found is performed before stacking and mosaicking the homogenised pawprints. PSF photometry is finally carried out on the deep, homogenised image. The procedure is documented in \citet{rubele2012,rubele2015}. 

The data analysed in this paper are from VMC tile LMC 6$\_$6, which has central coordinates of RA$_{\text{J}2000}$ = 05:37:40.008 and Dec$_{\text{J}2000}$ = $-$69:22:18.120. We use the deep $K_\text{s}$ tile. We choose the $\ks$ wavelength because it is minimally affected by extinction, but sill blue enough to be dominated by stellar sources rather than dust or nebular emission. The tile has been produced using 13 epochs of $K_\text{s}$ observations, with a total integration time of 9,315 s. We downloaded the deep tile, confidence map, and PSF catalogue from the VSA. We also downloaded the corresponding $Y$ and $J$ tiles and their confidence maps. The deep tiles were released with Data Release (DR) 6.\footnote{\url{https://archive.eso.org/cms/eso-archive-news/new-data-release-of-the-vista-magellanic-cloud-survey-vmc.html}} The deep tiles used in this analysis may be influenced by zero-point variations that may cause inhomogeneities on the deep images; these variations will be fixed in VMC DR7. 


\section{Overview of strategy}
\label{sec:data_reduction}

\begin{figure}
  \includegraphics[width=.95\linewidth]{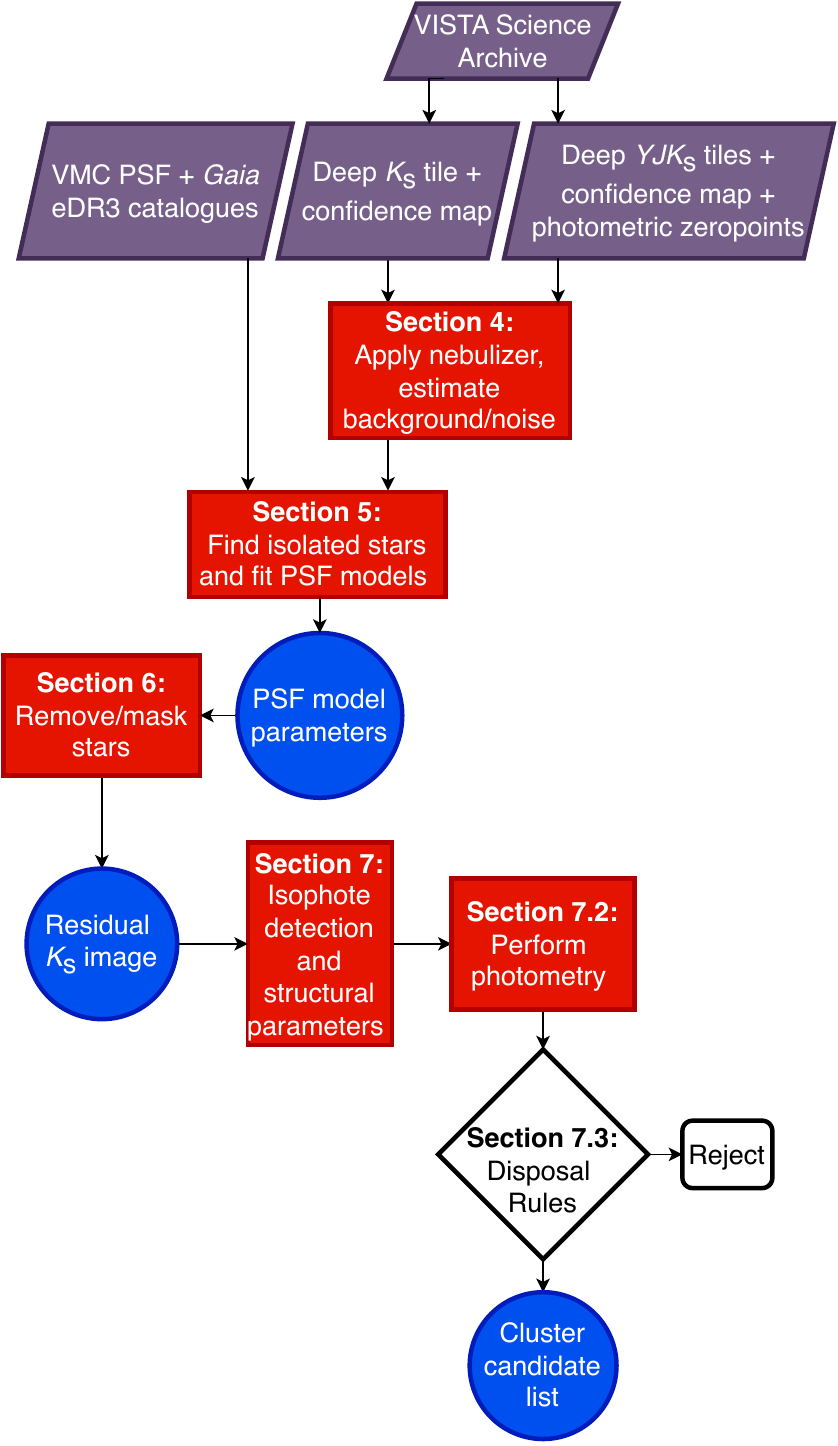}
  \caption{Schematic of our data processing steps, as more fully described in \S\ref{sec:bkg_noise}-\S\ref{sec:isophotes}. Processing steps are represented by red rectangles, external input as purple parallelograms, and input/output generated in the pipeline as blue circles. We note the relevant paper section in each step.}
  \label{fig:pipeline}
\end{figure}

Here we describe our method to remove point sources from deep, mosaicked $\ks$ images, with the goal of leaving extended objects like semi-resolved star clusters in the image. Figure \ref{fig:pipeline} is a schematic of our pipeline with the section of each step indicated in the diagram. 

Generally, it is cleanest to remove point sources from the single-exposure level \citep{montes2021}. We do not remove stars from the single-exposure level because the 5 s single-exposure $\ks$ images are not stored. The lowest-level $\ks$ images that are stored at the VSA are 75 s `normals' and 375 s pawprints. Here, we are using the terminology adopted on the VSA webpage.\footnote{\url{http://vsa.roe.ac.uk/dboverview.html}} We performed tests on the normals, pawprints, and deep pawprints, where deep pawprints are the result of stacking pawprints taken at different epochs. We found removing bright stars to be problematic on all images. Consequently, we devised a method that combines pixel-masking and PSF modeling and subtraction to cleanly remove bright stars. Upon refining this approach, we observed satisfactory results when applied to both deep and mosaicked images (tiles). To avoid extra computational time and significant residuals from stacking/mosaicking shallower images, we used our method on the deep VMC tile. 

Our method first models the background, finds isolated stars, models the PSF of the isolated stars, and then removes all stars from the deep $\ks$ VMC image. Finally, our pipeline detects extended objects on the image and calculates their properties. Our method is designed to deal with the following issues:
\begin{enumerate}[\bfseries (i)]
    \item An image with nebular emission that causes confusion between source and background.
    \item A PSF that varies across the image plane.
    \item A different PSF for bright and faint stars.
    \item An extremely crowded image in terms of number density of stellar sources.
    \item Detecting extended, semi-resolved objects on an image and splitting multi-peaked ones into multiple, hierarchical detections. 
    \item Culling spurious detections from the candidate list, like galaxies and blends.
\end{enumerate}
We describe in detail how we deal with these issues in each section and subsection below. 

\section{Background and noise modelling}
\label{sec:bkg_noise}
The deep $\ks$ tile considered in this paper contains many stellar sources and nebulous structures. The stars and nebulous structures have emission in $\ks$. The tile also has a background gradient and image artefacts. In order to separate stellar sources from this nebular emission and remove the background gradient, we first apply a background filtering routine to the deep $K_\text{s}$ tile by employing CASU's {\sc nebuliser} software.\footnote{\url{http://casu.ast.cam.ac.uk/surveys-projects/software-release/background-filtering}} This effectively removes nebulous structures that are non-stellar from the image, as well as any background gradient. We use the nebuliser software on the image, and then use the nebulised image to construct, model, and subtract the background using the following steps:

\begin{enumerate}[\bfseries (i)]
    \item We apply the nebuliser software with a sliding median filter box of 90 pixels and a sliding linear filter box of 30 pixels, consistent with CASU's recommendation of a 3-to-1 ratio.
    \item We use confidence maps from CASU to mask hot pixels, dead pixels, and bad pixels.
    \item We mask all pixels with values outside twice the standard deviation of the image.
    \item We find the median noise value and standard deviation in grids of $50 \times 50$ pixels.
    \item We perform sigma clipping and then apply a median filter with a size of three in each grid dimension (here, each grid dimension is a $50 \times 50$-pixel section).
    \item We create the background arrays and standard deviation arrays by interpolating across all pixels in the detector image.
    \item We then subtract the background in each grid region.
\end{enumerate}

Steps {\bf (iii)} through {\bf (vii)} were performed using tools within the \texttt{python} package \textsc{photutils}. 

\section{Modeling the VISTA deep tile PSF}
\label{sec:modeling_psf}
Here, we outline our approach to modeling the PSF of point sources in the nebulised, background-subtracted deep $\ks$ tile. Initially, we employ isophote contours to automatically detect isolated stars, organizing them into two groups based on their magnitude: faint and bright. We justify this division, highlighting the physical reasons why different PSF models are required for the PSF modeling. Subsequently, we display the analytical form of the PSF model used for the modeling process. Finally, we delineate the methodology for modeling both faint and bright sources' PSFs.

\subsection{Finding isolated stars}
\label{sec:isolated_stars}

\begin{figure}
  \includegraphics[width=.95\linewidth]{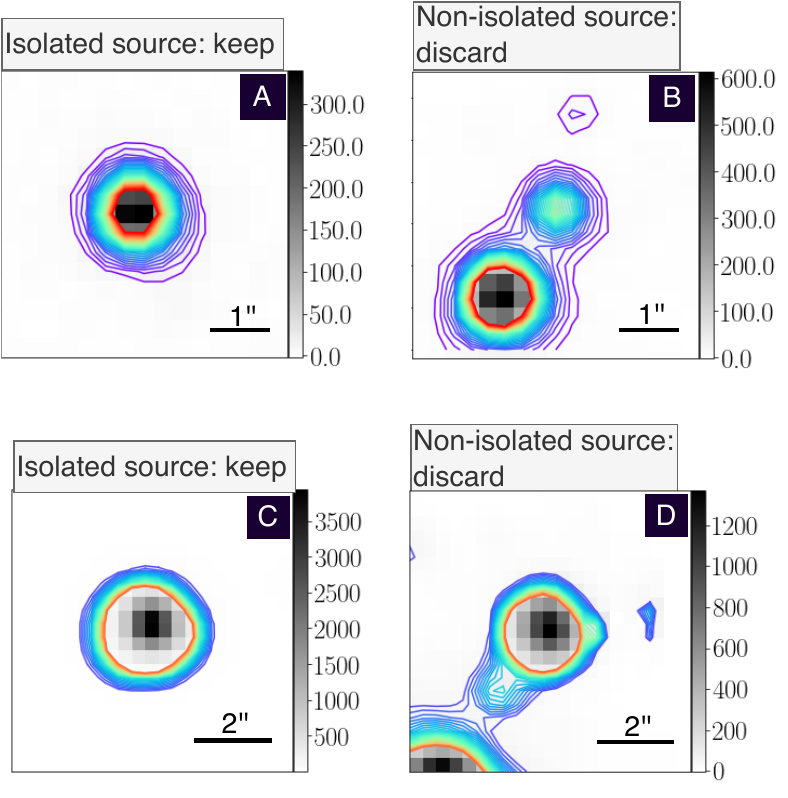}
  \caption{Demonstration of how we automatically select isolated stars; a description of this process is in \S\ref{sec:isolated_stars}. Panels A and B are stars in the faint magnitude bin, and panels C and D are for the bright magnitude bin. The faint and bright magnitude bins are described in the first paragraph of \S\ref{sec:isolated_stars}. Objects in panels A and C are classified as isolated during this procedure, and objects in panels B and D are classified as non-isolated and discarded.}
  \label{fig:isolated_stars}
\end{figure}

The next part of our method involves automatically selecting isolated point sources from the VMC PSF source catalogue. Selecting isolated sources is an important step in fitting PSF models. We do this in two magnitude bins: (1) `faint' stars with $\ks$ magnitudes between 14.5 and 18.0, and (2) `bright' stars with $\ks$ magnitudes between 9.0 and 15.5. We separate our stars into these two bins because we find that they are not well-fitted by identical analytic PSF models; however, there is overlap between the two categories, which is why there is overlap between our binning scheme. We follow different procedures in each bin for PSF model fitting (in \S\ref{sec:PSF_components}, we explain in detail why we do this). These are the steps we follow to find isolated point sources:

\begin{enumerate}[\bfseries (i)]
    \item Select faint or bright sources from the VMC PSF catalogue using cuts in stellar probability, $\tt{Star\_prob} > 0.5$, and the $K_\text{s}$ sharpness index, $\tt{SHARP} < 0.5$.\footnote{The sharpness index is the ratio of the fitted PSF FWHM to the model PSF FWHM, so extended objects have larger sharpness values. The stellar probability parameter is determined based on an object's locus in colour--colour space, the local completeness, and the sharpness index \citep[][further discusses the sharpness and stellar probability parameters]{bell2019}.}
    \item Use the background-subtracted image divided by the noise image to put the cutout in units of $\sigma$; then produce a cutout around the star.
    \begin{enumerate}[\bfseries a)]
        \item For the faint bin, the size of the cutout is $17 \times 17$ pixels (panels A and B in Figure \ref{fig:isolated_stars} show cutouts of stars in this magnitude bin).
        \item For the bright bin, the size of the cutout is $21 \times 21$ pixels, larger than the faint-star cutouts because bright stars have a larger skirt of light (panels C and D in Figure \ref{fig:isolated_stars} show cutouts of stars in this magnitude bin).
    \end{enumerate}
    \item Fit isophotes, using \textsc{scikit-measure} routines, to the sources at intermediate $\sigma$ thresholds.
    \begin{enumerate}[\bfseries a)]
        \item For the faint bin, the thresholds run from of 5$\sigma$ to 210$\sigma$ in increments of 5$\sigma$ (panels A and B in Figure \ref{fig:isolated_stars} show cutouts of stars in this magnitude bin with the isophotes shown as coloured contours).
        \item For the faint bin, the thresholds range from of 20$\sigma$ to 210$\sigma$ in increments of 5$\sigma$. A higher threshold starting point here implies that the isolated bright stars we select often have some faint neighbours, but we find that this does not affect our fits (panels C and D in Figure \ref{fig:isolated_stars} show cutouts of stars in this magnitude bin with the isophotes shown as coloured contours).
    \end{enumerate}
    \item If more than one source peak results from this procedure (in the form of more than one isophote at any $\sigma$ level), then the source is classified as non-isolated and removed (panels A and C of Figure \ref{fig:isolated_stars} show examples of isolated stars we keep, whereas panels B and D are examples of discarded sources). If the detected source's centroid is located at or beyond 0.4 pixels from the cutout centre, we discard the source because it is likely blended.
\end{enumerate}

After applying this procedure, there are approximately 55,000 isolated faint point-sources we detect and approximately 18,000 isolated bright point-sources. In the next sections, we fit PSF models to the isolated point-sources. 

\subsection{Why the PSF varies for stars of different magnitudes}
\label{sec:PSF_components}
In this section, we briefly explain why we separate the point sources into two different magnitude bins and apply a different PSF modelling procedure to each bin. The faint stars cover a magnitude range generally used to quantify `PSF stars \citep{jarvis2021}. PSF stars are used to model the shape of the core or inner PSF to be used for PSF photometry in software packages such as DAOPHOT \citep{Stetson1987}. For these purposes, the core PSF is generally described by the inner 0.5 to 2.0 arcsec \citep{Trujillo2001a,Trujillo2001b}. For ground-based telescope systems, the inner PSF shape is produced by atmospheric turbulence \citep{Racine1996}. We have found that stars within our faint magnitude range are well-fitted and well-removed by our analytic PSF models. The brighter point sources are not well-fitted and thus cannot be cleanly subtracted using the same analytic PSF models as used for the faint point sources. This ill-fitting emerges for three main reasons: first, the brighter--fatter effect, second, the outer PSF becoming more significant; and, third, the smudging of the centre of the PSF caused by stacking and mosaicking. We further detail these reasons below.
\begin{enumerate}[\bfseries (i)]
    \item The brighter--fatter effect 
\citep{Guyonnet2015,jarvis2021,Antilogus2014} is a nonlinear effect where a bright point source appears to have a larger size (FWHM) compared to that of a fainter point source. In optical CCDs, it is caused by accumulated charge in a pixel that changes the electric field's geometry: pixels receiving high flux repel new incoming electrons arriving later, causing them to fall into outer pixels. This effect is neither well-studied nor quantified in near-infrared array detectors, but it is still evident, albeit less severe \citep{Hirata2020}. This causes the core PSF width for bright stars to appear larger. 
\\
\item For brighter stars, the integrated light in the extended PSF wings starts to become significant \citep{Sandin2014}. The inner PSF generally used to quantify the PSF is a very small part of the full PSF; the full size of the extended PSFs can range up to and become even greater than 1 arcmin. Extended PSFs have complex structures that depend on the unique light path and its interaction with the atmosphere and all the elements of the telescope \citep{Infante-Sainz2020}. The outer PSF is sometimes called the `aureole' \citep{Racine1996, Sandin2014}. The shape of the aureole is not described by the same profile as that of the inner PSF and is not caused by atmospheric turbulence; it is thought to be caused by effects such as atmospheric diffusion, scattering of light by the telescope/instrument/detector optics, micro-ripples and dust on optical surfaces, and/or reflection within the instrument \citep{Racine1996,Sandin2014}. Studies have shown that the wings of an empirically generated PSF model have higher values compared to those of a fitted analytical model \citep{montes2021, 2022MNRAS.513.1459M}. Since the wings are negligible for faint stars but important for bright stars, faint and bright stars are not well-fitted by identical analytical models.
\\
\item The deep tile we are using is generated by stacking and mosaicking $\sim$390 images, each with an exposure time of 75 s. These $\sim$390 images themselves are the product of stacking up to 15 single-exposure 5 s images. The stacking and mosaicking process alters the shape of the PSF. More importantly, the stacking and mosaicking causes smudging of the centre of the PSF, which causes the maximum-intensity pixels to spread amongst many pixels, instead of just one. This distorts the appearance of the PSF in the deep, mosaicked image. This occurs for all point sources but only becomes noticeable for bright ones. In turn, this causes large residuals at the centre of the source when we attempt removal with an analytic PSF model.
\end{enumerate}

To correct for {\bf (i)}, studies often remove brighter stars from their PSF star sample or apply a correction to reassign flux in an image back to its original position \citep{jarvis2021}. To deal with  {\bf (ii)}, considerable effort has been undertaken to model the various parts of the PSF, depending on the science case. For example, \citet{montes2021} created empirical models of the core, the intermediate, and the outer parts of the PSF in order to remove them from optical images, with the goal of studying low-surface-brightness galaxies. To deal with  {\bf (iii)}, studies that remove stars from images remove them at the single-exposure level \citep{montes2021}. 

To deal with these issues in the context of our method, we have adopted a different strategy to modelling the PSF of brighter stars. For the stars in the brighter magnitude bin, we mask out the highest-intensity pixels corresponding to the source. This effectively masks out the core or inner part of the PSF. Then, we fit the analytic PSF models to the outer PSF. This deals with the inner PSFs being larger for bright than for faint stars, the outer PSFs having a different profile from the inner PSF, and the smudging of the centre of the PSF. 

\subsection{Analytic PSF models}
\label{sec:analytic_psf_models}

At various stages in the next sections which describe the next steps in our pipeline, we will fit PSF models to point-source cutouts. We adopt three functional forms of analytic PSF models for this purpose: a rotated, elliptical Gaussian model (six free parameters), a rotated, elliptical Moffat profile (seven free parameters), and a compound model consisting of a sum of Gaussian and Moffat components (11 free parameters). The form of the last and most general of these is: 
\begin{align}
\begin{split}
f(x,y) &= \begin{aligned}[t]
A [ 1 + a(x-x_0)^2 + 2b(x-x_0)(y-y_0), \\
+ \text{ } c(y-y_0)^2]^{-\beta} + A^{\prime} \exp (- (a^{\prime}(x-x_0)^2,   \\
 + \text{ }  2b^{\prime}(x-x_0)(y-y_0) + c^{\prime}(y-y_0)^2)),\\
a = \frac{\cos^2 \theta}{\alpha_x^2} + \frac{\sin^2 \theta}{\alpha_y^2}, \\
b = \frac{\sin 2\theta}{2\alpha_x^2} - \frac{\sin 2\theta}{2\alpha_y^2}, \\
c = \frac{\sin^2 \theta}{\alpha_x^2} + \frac{\cos^2 \theta}{\alpha_y^2}, \\
a^{\prime} = \frac{\cos^2 \theta^{\prime}}{2\sigma_x^2} + \frac{\sin^2 \theta^{\prime}}{2\sigma_y^2}, \\
b^{\prime} = \frac{\sin 2\theta^{\prime}}{4\sigma_x^2} - \frac{\sin 2\theta^{\prime}}{4\sigma_y^2}, \\
c^{\prime} = \frac{\sin^2 \theta^{\prime}}{2\sigma_x^2} + \frac{\cos^2 \theta^{\prime}}{2\sigma_y^2} .
\label{eqn:psf_model}
\end{aligned}
\end{split}
\end{align}
In this expression, $x_0$ and $y_0$ are the $x$ and $y$ centres, $A$ and $A'$ are the amplitudes of the Moffat and Gaussian components, respectively, $\theta$ and $\theta'$ are the rotation angles of these two components (defined as positive anticlockwise from the major axis, in the range $-90$ degrees to $+90$ degrees), $\sigma_x$ and $\sigma_y$ are the standard deviations of the Gaussian component along the major and minor axes prior to rotation, and $\alpha_x$ and $\alpha_y$ are the core width along the major and minor axes before rotation, respectively. To fit a purely Gaussian or Moffat model, we simply set $A=0$ or $A' = 0$, respectively, in the expression above; we refer to models in which neither $A$ nor $A'$ is set to zero as Compound models. In practice, to fit purely Gaussian models we use the built-in Gaussian 2D \textsc{astropy} model, while we use custom \textsc{astropy} models for the Moffat and Compound cases.

\subsection{PSF modelling approach for faint point sources}
\label{sec:faint_modeling}

\begin{figure*}
  \includegraphics[width=.8\linewidth]{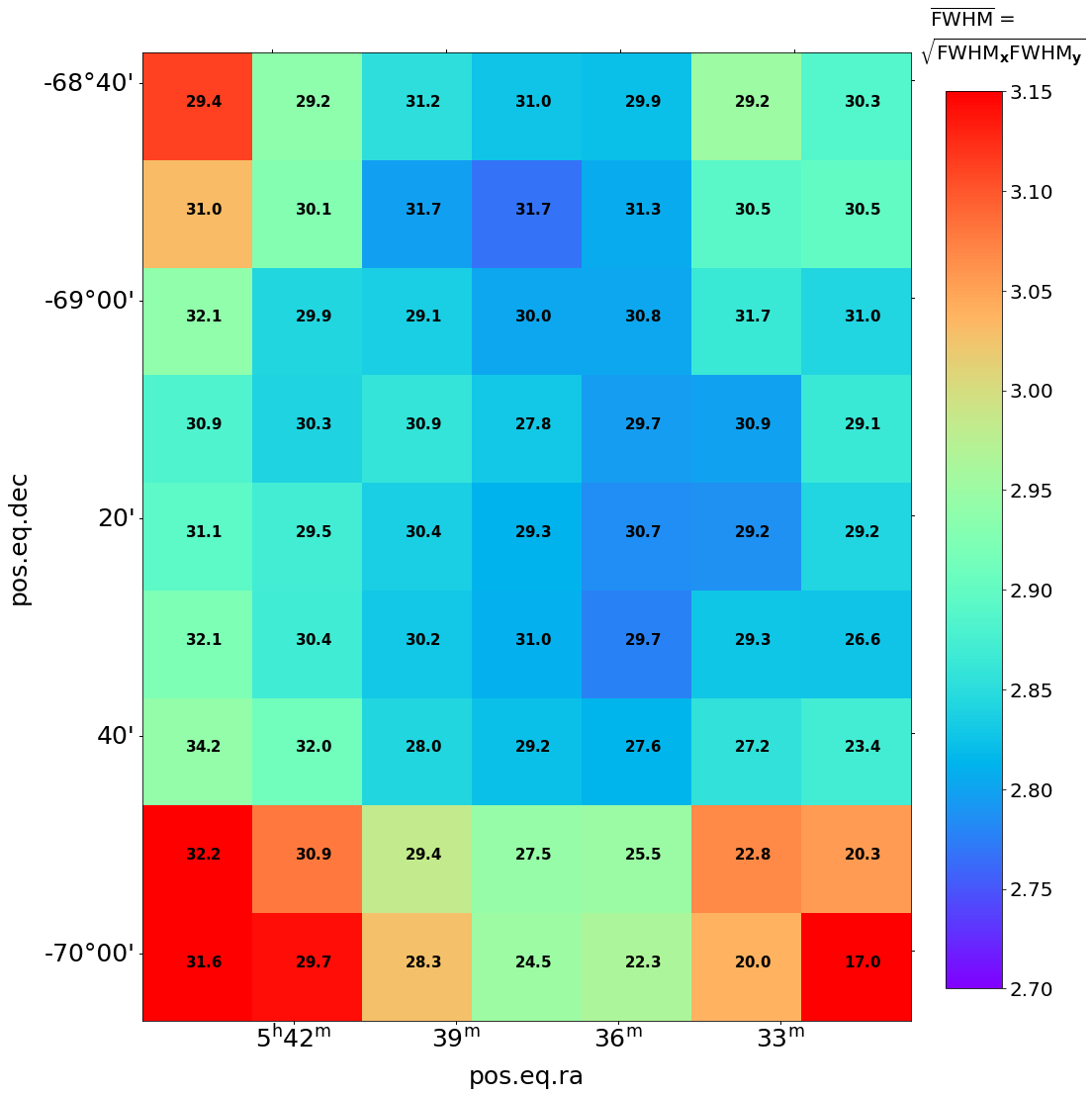}
  \caption{Here we show the mean effective PSF width from fitting 2D Gaussian models to isolated stars across the deep $\ks$ tile, dividing the tile into 56 grid cells. This procedure is documented in \S\ref{sec:faint_modeling}.  This image shows that the PSF size varies across the image, and so we must incorporate this effect when removing point sources. In each grid cell, we show show the square root of the number of stars in that cell because this is how the Poisson noise on the width, due to estimating it from a finite, discrete number of stars, would scale. The bottom right $\sim$4 bins are where VIRCAM detector 16 is located; known problems with this detector cause fewer stars to be detected on it. Other VIRCAM detectors have problems as well, but we do not see those reflected in a significant reduction in the number of observed stars. Another issue to note is that the left and right edges of this image have about half the exposure time of most of the tile; the affected area corresponds to roughly 15$\%$ of the tile. We do not seem to see a reduction in  isolated-source density due to this effect however. The problems with the VIRCAM detectors, and further details on a tile's variable depth, are given on the CASU webpages, \url{http://casu.ast.cam.ac.uk/surveys-projects/vista/technical/known-issues}.
  \label{fig:faint_psf}
  }
\end{figure*}

In this section, we describe how we fit PSF models to the faint point sources. Our strategy is to fit three PSF models to every faint point source we detected in \S\ref{sec:isolated_stars}, and then save the best-fitting model parameters and location in the image. We do this because the parameters of the PSF vary significantly across the image. To show this, we refer to Figure \ref{fig:faint_psf}; this figure is the result of applying the analysis procedures of this section to the 54,000 isolated, faint point sources we found in \S\ref{sec:isolated_stars}. In Figure \ref{fig:faint_psf} we show the mean effective PSF width we obtain. The PSF width varies by about 20\%. The PSF variation of VISTA images is caused predominantly by optical distortions of the telescope system \citep{sutherland2015}; further, the stacking and mosaicking of VISTA images also cause variations in the PSF. In one VISTA tile image, there can be over 100 PSF forms \citep{gonzalez2017}. This demonstrates that we cannot use one `faint' PSF model to remove all stars in the faint magnitude bin. 

\begin{figure*}  
\includegraphics[width=.85\linewidth]{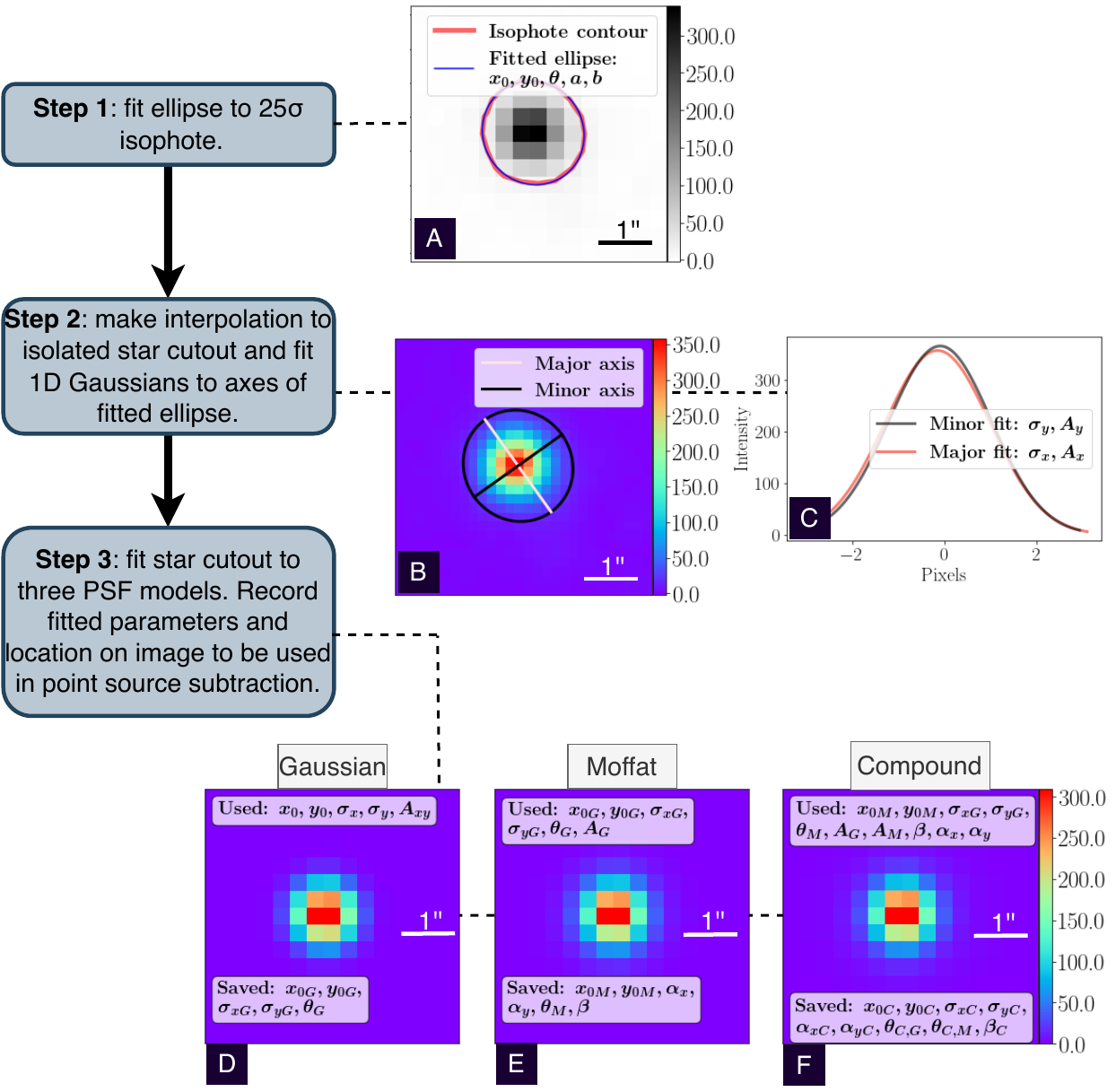}
  \caption{This figure shows how we fit PSF models to stars in our faint ($\ks$ magnitude 14.5 to 18.0) magnitude bin. These Steps are described in detail in \S\ref{sec:faint_modeling}. Step 1 and panel A shows
  fitting an ellipse to an isolated star at 25$\sigma$. Step 2 and panels B and C show interpolating the star cutout and using it to fit 1D Gaussians along major/minor axes. Step 3 and panels D, E and F show using the ellipse and 1D Gaussian parameters to fit a 2D rotated Gaussian, using the 2D Gaussian parameters to fit a 2D rotated Moffat, and then using both the 2D Gaussian and 2D Moffat parameters to fit a 2D rotated Compound (Gaussian + Moffat). Equation \ref{eqn:psf_model} describe the Gaussian, Moffat, and Compound model we use. In panel D, $x_0$ and $y_0$ are centre coordinates from ellipse fitting; $\sigma_x$ and $\sigma_y$ are the $x$ and $y$ standard deviations from the 1D Gaussian fitting and $A_{xy}$ is the mean of the fitted 1D Gaussian amplitudes, $A_x$ and $A_y$. In panels D, E, and F, the remaining parameters listed are the same as in equation \ref{eqn:psf_model}; the subscripts $G$, $M$, and $C$ indicate a best-fit parameter from the 2D Gaussian, Moffat, or Compound, respectively. In the bottom three plots (D, E and F), the parameters labelled `Used' are used as initial guesses for the fitter and the parameters labelled `Saved' are saved parameters to be used later in source subtraction; source subtraction is described in \S\ref{sec:star_remove}. All colour bars are in units of background-subtracted counts. }
  \label{fig:faint_star_models}
\end{figure*}

Figure \ref{fig:faint_star_models} shows the processes of fitting PSF models to isolated, faint point sources. The steps we follow are:

\begin{itemize}
\item[{\bf Step 1:}] Fit an isophote to the sources' pixels at a threshold of 25$\sigma$ and then fit an ellipse to the isophote vertices (this is shown in panel A in Figure \ref{fig:faint_star_models}). The isophotes and ellipses are fitted using \textsc{scikit-image measure} routines. The ellipse is defined by its centre, major axis along the $x$ axis before rotation by the position angle, minor axis along the $y$ axis before rotation by the position angle, and the position angle. The position angle is converted to a range between $-90$ degrees and $+90$ degrees.\\
\item [{\bf Step 2:}] Interpolate the cutout using a 2D--Tocher interpolater, implemented with \textsc{scipy}, to obtain interpolated pixel values along the major and minor axes of the best-fitting ellipse (shown in panel B of Figure \ref{fig:faint_star_models}). Then, fit 1D Gaussians along the major and minor ellipse axes (shown in panel C in Figure \ref{fig:faint_star_models}). The fitting routine returns the amplitude, centre and standard deviation along each axis of the ellipse. \\
\item [{\bf Step 3:}] Fit three PSF models to the isolated stars. We use the Levenberg--Marquardt algorithm, implemented by \textsc{scipy}; it is a non-linear least-squares algorithm, and performs best when given a reasonable initial guess for the relevant model parameters. We use the ellipse and 1D Gaussian parameters as initial guesses for the 2D Gaussian, then the best-fitting parameters from the 2D Gaussian as initial guesses for the 2D Moffat, and then the best-fitting parameters from the Gaussian and Moffat as initial guesses for the Compound model (shown in panels D, E and F in Figure \ref{fig:faint_star_models}). We save the locations of the stars and their PSF model parameters. 
\end{itemize}

In order to save the parameters later for point-source subtraction, we divide the image into 56 grid cells, where each cell has a size slightly smaller than 2,000 $\times$ 2,000 pixels (Figure \ref{fig:faint_psf} shows how we grid sources in the image). This size is chosen because it is comparable to the size of a VIRCAM detector. Within each grid cell, we record the median and standard deviation of each of the parameters obtained from fitting the PSF of each of the stars in the bin.

\subsection{PSF modelling approach for bright point sources}
\label{sec:bright_modeling}

\begin{figure*}
  \includegraphics[width=.95\linewidth]{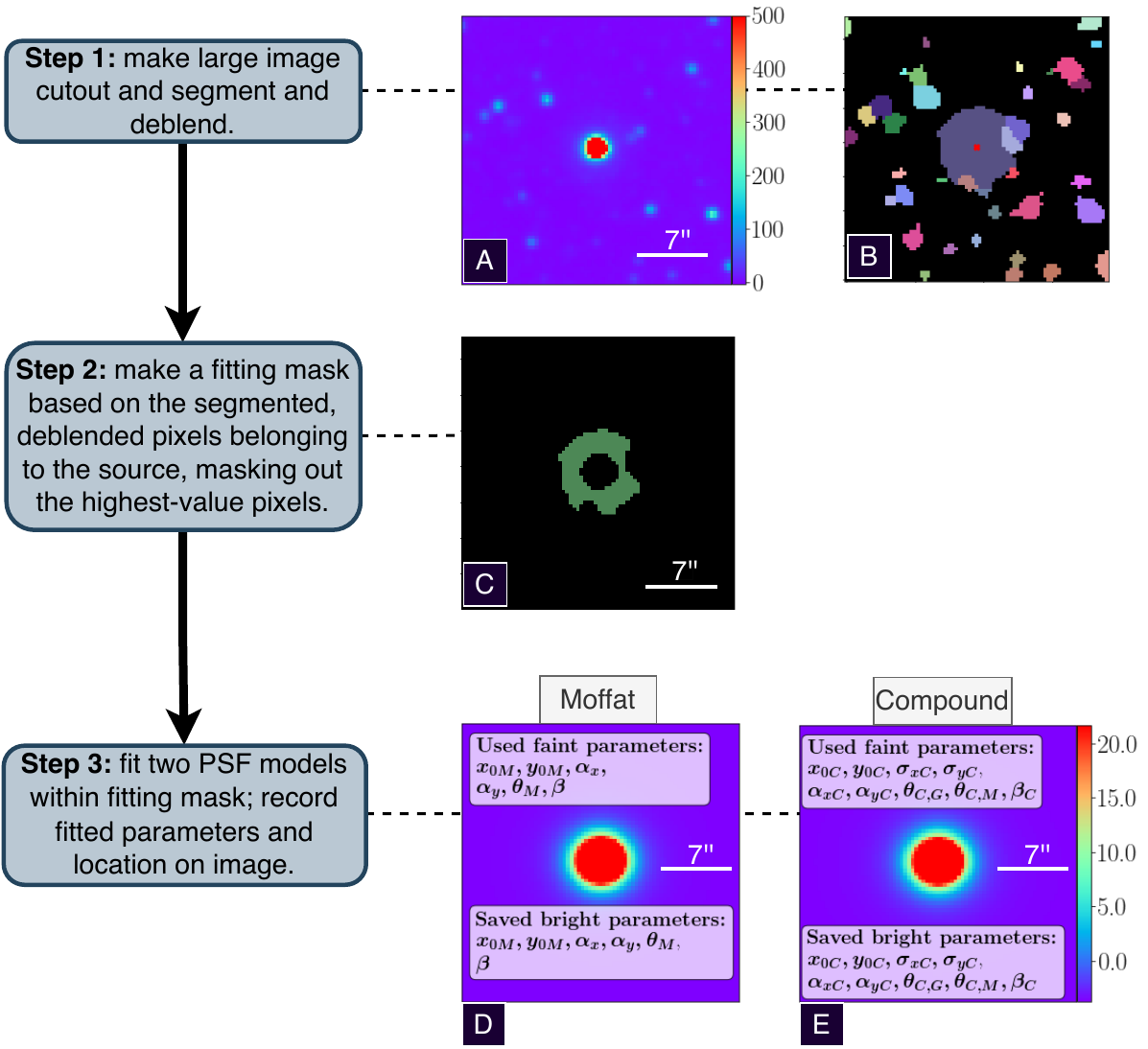}
  \caption{The process of fitting PSF models to isolated, bright ($\ks$ magnitude 9.0 to 15.5; \S\ref{sec:isolated_stars} describes the faint and bright magnitude bins used) point sources. Step 1 is making a large cutout and then segmenting and deblending the cutout (shown in panels A and B). Step 2 is making a fitting mask for the outer PSF and masking brightest pixels (an example of such fitting mask is shown in panel C). Step 3 is fitting two PSF models to the outer PSF mask (shown in plots D and E). In plots D and E, the `used faint parameters' are the PSF model parameters from \S\ref{sec:faint_modeling} in the same region of the image as the source being modelled (the image is split into 72 grid cells, each approximately 2,000 $\times$ 2,000 pixels; the grid cells are shown in Figure \ref{fig:faint_psf}). All colour bars are in units of background-subtracted counts. The `saved bright parameters' are the best-fitting Moffat and Compound parameters saved during the fitting process; the caption of Figure \ref{fig:faint_star_models} describes this. The main different between the procedure for the faint point sources, which was outlined in Figure \ref{fig:faint_star_models}, is that here we fit the PSF models only to the pixels belonging to the outer PSF.}
  \label{fig:bright_star_models}
\end{figure*}

Figure \ref{fig:bright_star_models} displays a diagram of our procedure for selecting isolated, bright stars and fitting PSF models. We apply this analysis to the 18,000 isolated, bright point sources we detected in \S\ref{sec:isolated_stars}. The steps we follow are: 

\begin{itemize}
    \item [{\bf Step 1:}] Make a $77 \times 77$ pixel cutout around the source (panel A in Figure \ref{fig:bright_star_models} shows such a cutout). We use such a large cutout because some brighter stars have very large skirts of light. Segment the cutout using \textsc{photutils.segmentation} tools. First, we detect groups of five pixels at a threshold of 3$\sigma$. Then, we deblend the detections with minimum numbers of connected pixels of 50, and thresholds at 1,000 levels with the contrast set to zero and the mode set to exponential. This means that, at each thresholded level, all peaks are considered separate sources. Using the segmented, deblended star cutout, we can select the pixels belonging to just the source and exclude those belonging to any of its neighbours. In panel B in Figure \ref{fig:bright_star_models}, the purple pixels at the centre belong to the bright source to which we will fit the PSF model. \\
    
    \item [{\bf Step 2:}] Create a fitting mask to fit the outer PSF. We do this by masking the highest-intensity pixels among those assigned to the source during the segmentation step. Based on several trial values, we set the mask level to a number of background-subtracted counts equal to the total integration time in seconds divided by 65. This number is driven purely by experimentation. Panel C in Figure \ref{fig:bright_star_models} shows an example of a fitting mask (the green pixels) created using this procedure. \\
    
    \item [{\bf Step 3:}]  Fit two PSF models (the Moffat and Compound; we do not fit the bright stars with the Gaussian model because it has the worst performance for bright point sources) to the outer PSF fitting mask and record the parameters and locations on the image (panels D and E show examples of best-fitting Moffat and Compound models in Figure \ref{fig:bright_star_models}. As initial guesses for the Levenberg--Marquardt fitter, we use the saved PSF parameters from \S\ref{sec:faint_modeling} at the same location in the image as the source we are fitting.
\end{itemize}

After applying this process to all bright stars across the image, we save the locations and PSF parameters of approximately 18,000 isolated, bright point sources. We use the same procedure as for the faint stars.

\section{Creation of the residual image}
\label{sec:creation_residual_image}
Here, we describe how we remove stars based on their catalogue magnitude. Then, we present the residual image. 

\subsection{Removing point sources}
\label{sec:star_remove}

\begin{figure*}
  \includegraphics[width=.95\linewidth]{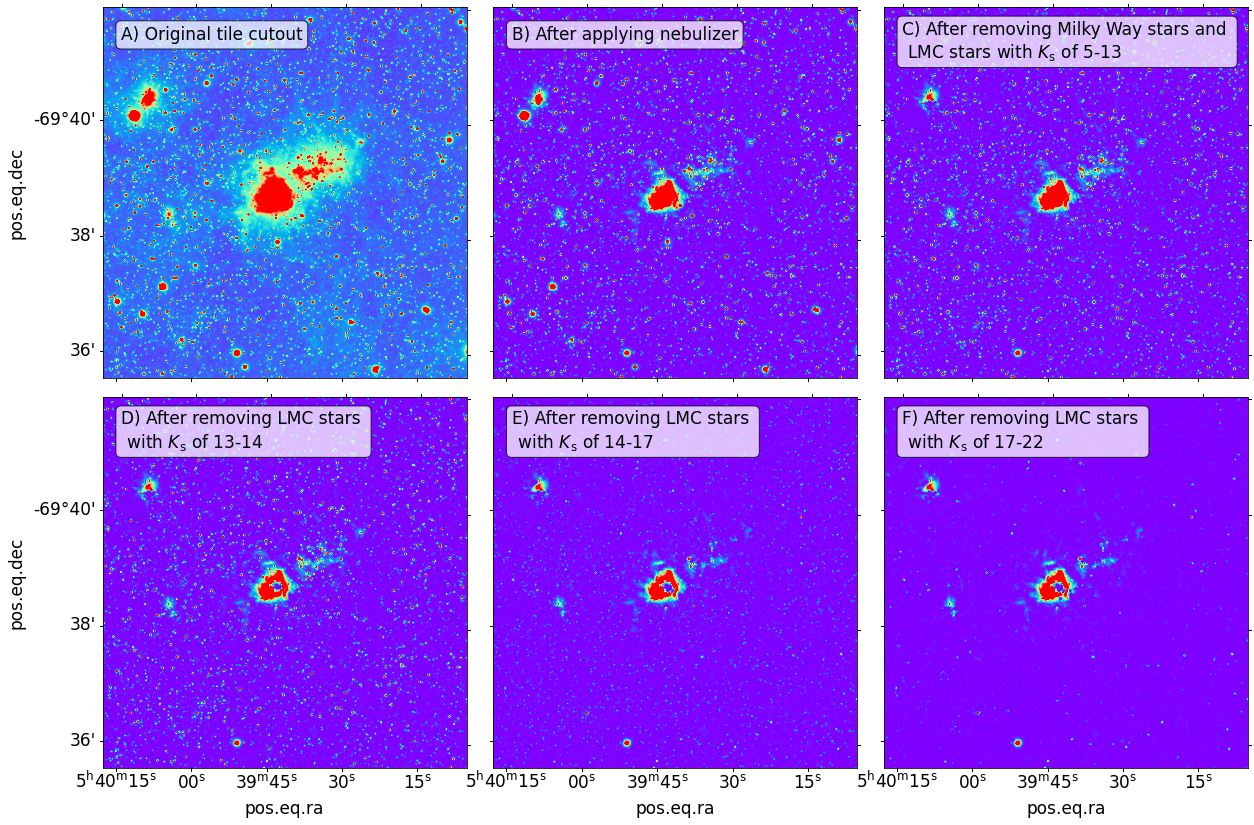}
  \caption{This image shows each stage of image reduction and point source removal of the deep $\ks$ tile around one 1,000 $\times$ 1,000 pixel cutout. Panel A shows the original tile after downloading from the VSA; panel B shows the cutout after applying the nebuliser software; and panels C through F show the cutouts after removing stars in each magnitude bin. Each stage of source-removal is described in \S\ref{sec:creation_residual_image}.}
  \label{fig:progress}
\end{figure*}

\begin{figure*}
  \includegraphics[width=.85\linewidth]{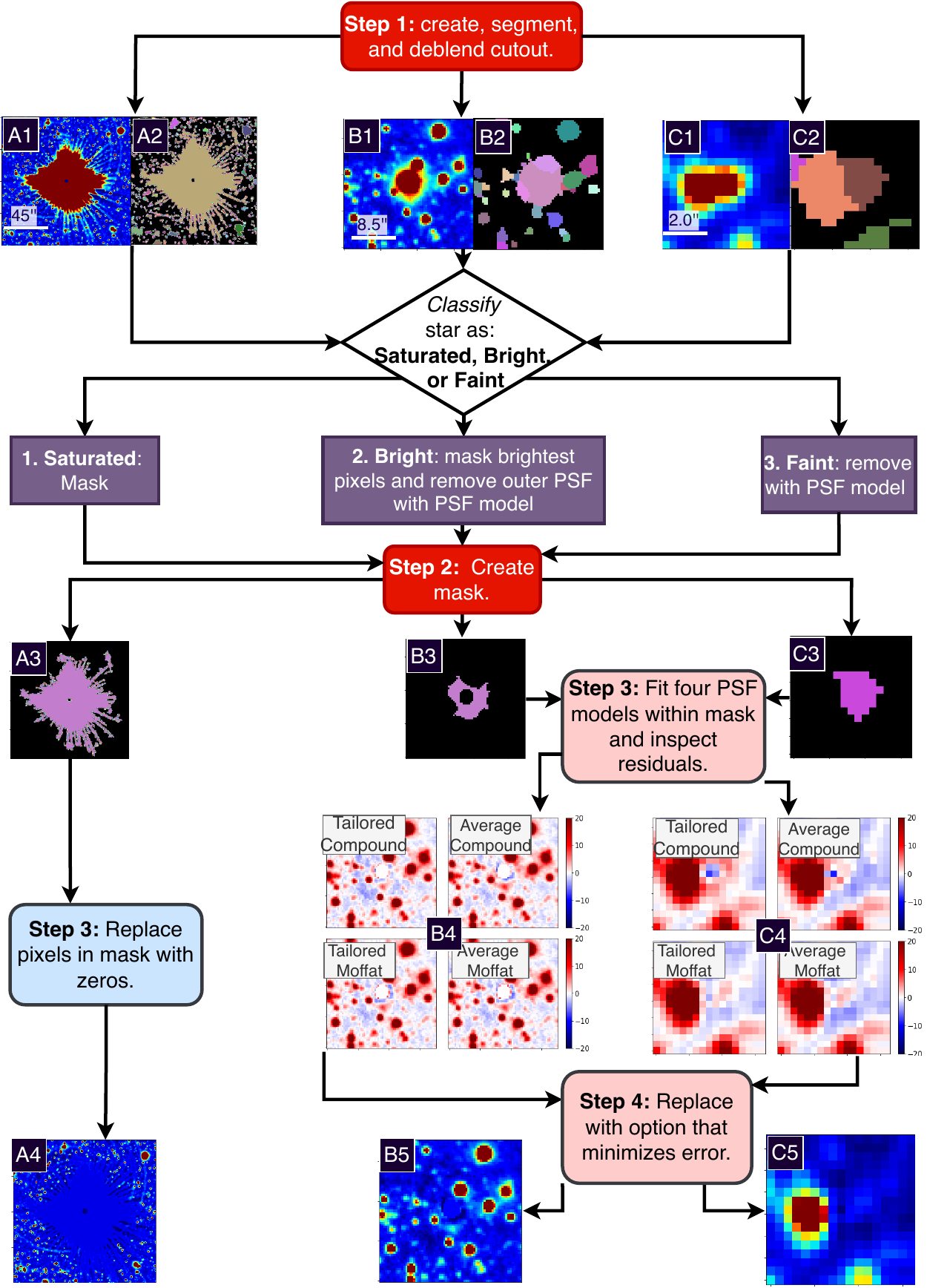}
  \caption{Diagram showing the process of star removal. In Step 1, the size of the cutout and the method of segmentation and deblending is driven by the catalogue magnitude of the source. The brightest stars, like those shown in the panels labelled A, are given the largest cutouts and subjected to less deblending. The next brightest are given smaller cutouts and the deblending is increased, as in the panels labelled B. The faintest stars have the smallest cutouts and maximum deblending; we show this in the panels labelled C. After segmentation, the stars are classified as saturated, bright or faint, based on a magnitude calculated in our pipeline and properties of the pixels belonging to the source in the segmented image. Then, depending on the classification, a fitting mask is created and the star is removed.}
 \label{fig:remove_stars}
\end{figure*}

The next step in our pipeline is to remove point-like sources using our PSF model parameters. First, we apply the nebuliser software again to the raw, deep tile image with a median filter size of 30 pixels and a linear filter size of 10 pixels (in \S\ref{sec:bkg_noise}, we describe the parameters of the nebuliser software). We then estimate the background and noise in the image using the same procedure as outlined in \S\ref{sec:bkg_noise}. Applying the nebuliser software more aggressively at this stage helps to remove all nebulous, non-stellar emission from the image. We applied it less aggressively in the previous sections when fitting the PSF models to avoid removing extended emission from the brightest stars, which we need to keep in order to fit their PSFs accurately. This extended emission is no longer needed once we have the PSF model in hand. Figure \ref{fig:progress} panel A shows a cutout of the raw tile and panel B shows the cutout after applying the nebuliser. 

We remove stars iteratively from the nebulised, background-subtracted deep tile in the following order:

\begin{enumerate}[\bfseries (i)]
    \item Milky Way foreground point sources using a cross-match between VMC and {\sl Gaia} early Data Release 3 \citep[EDR3;][]{gaiaedr3}. We select stars that have parallax values $>$ 0.2 mas and parallax errors $<$ 0.1 mas.
    \item Point sources (selected with $K_\text{s}$ $\tt{SHARP}$ index $<$ 0.5 and $\tt{Star\_prob}$ $>$ 0.33) from the VMC PSF catalogue in four $\ks$ magnitude bins:
    \begin{enumerate}[\bfseries a)]
        \item 5 to 13 mag (shown in panel C of Figure \ref{fig:progress});
        \item 13 to 14 mag (shown in panel D of Figure \ref{fig:progress});
        \item 14 to 17 mag (shown in panel E of Figure \ref{fig:progress});  
        \item 17 to 22 mag (shown in panel F of Figure \ref{fig:progress}).
    \end{enumerate}
\end{enumerate}

Within each magnitude bin, the same basic procedure is applied. First, we produce a cutout around the source in the background-subtracted and noise images. It is very important that the cutout is large enough and encompasses the extended PSF; therefore, we make larger cutouts for brighter stars and smaller cutouts for fainter stars. Table \ref{tab:star_stats} gives the size of the cutout we use for each magnitude bin. 

\begin{table}
    \caption{Table showing the point-source type, the number of sources of that type, and the size of the cutout we make around each source in order to remove it. The point-source type and how we remove them are described in \S\ref{sec:star_remove}.}
    \begin{tabular}{lcc}
        \hline
        \textbf{Point-source} & \textbf{Number of} & \textbf{Cutout size} \\
        \textbf{type} & \textbf{ sources} & \textbf{(pixels)} \\
        \hline
        Milky Way ({\sl Gaia}/VMC) & 790 & 401 $\times$ 401 \\
        LMC: 5 -- 13 mag (VMC) & 7,840 & 101 $\times$ 101 \\
        LMC: 13 -- 14 mag (VMC) & 11,232 & 51 $\times 51$ \\
        LMC: 14 -- 17 mag (VMC) & 224,345 & 31 $\times$ 31 \\
        LMC: 17 -- 22 mag (VMC) & 804,461 & 17 $\times 17$ \\
        \hline
    \end{tabular}
    \label{tab:star_stats}
\end{table}

Next, we segment and deblend the cutout. The cutout and segmentation is shown as Step 1 in Figure \ref{fig:remove_stars}. Using the catalogue magnitude and properties of the segmented image, we then classify the star as saturated, bright, or faint. Saturated stars are identified because the saturated pixels are set to zero in the VDFS pipeline (\S\ref{sec:VMC} describes the VDFS pipeline). Most stars in the first four bins are classified as `bright.' Point sources are classified as `faint' and removed in a different way if there are not many pixels in the segmented image belonging to the source being removed. Finally, some of the point sources in the fourth magnitude bin and all of the point sources in the fifth magnitude bin are classified as `faint.' Each of the three classifications has a different removal procedure, which we describe in the next sections. A diagram showing a breakdown of each classification and how we remove the stars is shown in Figure \ref{fig:remove_stars}; panels labelled A show the procedure for saturated stars, panels labelled B show the procedure for bright stars, and panels C show the procedure for faint stars. As we iterate through the sources, we check to determine whether a source has already been removed, because there are duplicates and false detections in the PSF catalogue.

\subsubsection{Saturated source removal}
\label{sec:saturated_removal}

A saturated source is detected if one or more of the central pixels are negative, as the VDFS pipeline sets them to zero and they then become negative after sky subtraction. If a saturated source is detected, we follow the steps illustrated in Figure \ref{fig:remove_stars}. The steps we follow are: 

\begin{itemize}
    \item [{\bf Step 1:}] Make a cutout with size indicated in Table \ref{tab:star_stats} (panel A in Figure \ref{fig:remove_stars} shows this). Next, segment the cutout using \textsc{photutils.segmentation} tools. First we detect groups of 10 pixels at a threshold of 3$\sigma$. Next, deblend the segmented cutout based on the catalogue magnitude panel A2 in Figure \ref{fig:remove_stars} shows this): 
    \begin{enumerate}[\bfseries a)]
        \item Saturated stars with magnitude $<$8: deblend with minimum number of connected pixels\footnote{This number is a positive integer that describes the minimum number of connected pixels a source must have, each greater than the threshold value, in order to be deblended into a separate source.} of 50 and the contrast parameter set to 0.5. 
        \item Saturated stars with magnitude $>$8: deblend with minimum number of connected pixels of 20 and the contrast parameter set to 0.1. 
    \end{enumerate}
    \item [{\bf Step 2:}] Create a mask of pixels belonging to the source from the segmented, deblended image (shown in panel A3 in Figure \ref{fig:remove_stars}).\\
    \item [{\bf Step 3:}] Replace pixels in the mask with zeros (shown in panel A4 in Figure \ref{fig:remove_stars}). \\
\end{itemize}

We note that this procedure masks other sources that are close to the saturated source. Experimenting with fitting PSF models to the saturated source and removing them, we found that each saturated source required a tailored fitting procedure and so we opted to mask them instead.  

\subsubsection{Bright point-source removal}
\label{sec:bright_removal}

We next describe the bright point-source removal steps. Since the PSF varies so much across the image, the PSF varies for stars of different magnitudes, and the stacking/mosaicking of the image affects the shape of the PSF in unpredictable ways, our strategy is to fit PSF models to the outer PSF of each star individually and then remove the star with the option that minimises the error. We fit PSF models (with parameters described in detail in \S\ref{sec:analytic_psf_models}) with the Levenberg--Marquardt algorithm. As initial guesses and bounds for the fitting algorithm, we start by using the location of the point source in the image and figuring out which of the 56 bins it falls into (Figure \ref{fig:faint_psf} shows how we bin the image). Then, we use the median of the parameters in each bin as initial guesses and for the standard deviations to be used in the fitting bounds (\S\ref{sec:bright_modeling} describes how we bin the image and the parameters we save). 

We now describe our removal procedure in more detail, as it depends on magnitude. The steps for each magnitude group are shown in Figure \ref{fig:remove_stars}. The steps we follow are:

\begin{enumerate}[\bfseries (i)]
    \item \textbf{Non-saturated sources with $K_\text{s}$ $<$ 11 mag}:
    \begin{itemize}
    \item [{\bf Step 1:}] Create cutout with size indicated in Table \ref{tab:star_stats}. Then, detect and segment groups of at least 10 pixels in the cutout at a level of 3$\sigma$. Next, deblend the segmented cutout with the minimum number of pixels equal to 12 and contrast equal to zero.\\
        \item [{\bf Step 2:} ]Create a fitting mask by masking out the highest-intensity pixels in the sources' segmented pixels (using the threshold value given in \S\ref{sec:bright_modeling}).\\ 
        \item [{\bf Step 3:}] Fit two PSF models: 
        \begin{enumerate}[\bfseries 1)]
            \item A tailored Moffat, where $A$, $\alpha_x$, $\alpha_y$, $\beta$, and $\theta$ are allowed to vary with no bounds, and $x_0$ and $y_0$ are allowed to vary by 10 pixels from the centre of the cutout. 
            \item An average Moffat where $\alpha_x$, $\alpha_y$, and $\beta$ are held fixed, $A$ and $\theta$ are allowed to vary with no bounds and $x_0$ and $y_0$ are allowed to vary by 10 pixels.
        \end{enumerate}
        \item [{\bf Step 4:}] Remove the star with the option that minimises the error and set the brightest source pixels to zero. \\
    \end{itemize}
    \item \textbf{Sources with $K_\text{s}$  magnitudes between 11 and 13}:
    \begin{itemize}
        \item [{\bf Step 1:}] Create a cutout with size indicated in Table \ref{tab:star_stats} (panel B1 in Figure \ref{fig:remove_stars} shows an example of a cutout in this magnitude range). Next, detect and segment groups of 10 pixels in the cutout at a level of 3$\sigma$. Then, deblend the segmented cutout with the minimum number of pixels equal to one and contrast equal to zero (panels B2 in Figure \ref{fig:remove_stars} show a segmented, deblended cutout in this magnitude range).\\
        \item [{\bf Step 2:}] Create a fitting mask as described in Step 2 in (i). Check that the fitting mask has at least 40 pixels. If not, then remove it using the procedure outlined in \S\ref{sec:faint_removal} (panel B3 in Figure \ref{fig:remove_stars} shows a fitting mask in this magnitude bin). \\
        \item [{\bf Step 3:}] Fit four PSF models within the fitting mask: 
        \begin{enumerate}[\bfseries 1)]
            \item A tailored Moffat, where $A$, $\alpha_x$, $\alpha_y$, $\beta$, and $\theta$ are allowed to vary within 2$\sigma$ of the initial guesses, and $x_0$ and $y_0$ are allowed to vary by 4 pixels. 
            \item An average Moffat where all parameters are held fixed (the values being determined by the median value of the best-fitting bright PSF model resulting in the local image region) except for $A$, $\theta$, and the centre coordinates. 
            \item A tailored Compound model where $x_0$ and $y_0$ can vary within 4 pixels of the cutout's centre and all other parameters are allowed to vary within 2$\sigma$. 
            \item An average Compound model where all parameters are held fixed except for $A$, $A^{\prime}$, $\theta$, $\theta^{\prime}$, and the centre coordinates.\\
        \end{enumerate}
        \item [{\bf Step 4:}] Then, remove source using the model that minimises the least-squares error within a radius of 30 pixels from the best-fitting centre and set the brightest pixels to zero (panels B4 and B5 in Figure \ref{fig:remove_stars} show the residuals of removing a star in this magnitude range). \\
    \end{itemize}
    \item \textbf{Sources with $\ks$ magnitudes between 13 and 14:}
    \begin{itemize}
        \item [{\bf Steps 1 and 2:}] Segment, deblend, and create the fitting mask in the same manner as in Steps 1 and 2 of (ii). Check that the fitting mask has more than 40 pixels. If it has fewer, go to \S\ref{sec:faint_removal}.\\
        \item [{\bf Step 3:}] Fit the same four PSF models as in Step 3 of (ii), except that the centre can only vary by two pixels from the cutout's centre and the bounds of the parameters in the tailored Moffat and tailored Compound models are only allowed to vary within 1$\sigma$ from the initial guesses (the median and standard deviation of the model parameters in the local image area).\\
        \item [{\bf Step 4:}] Remove the source using the model that minimises the least-squares error within a radius of 15 to 20 pixels (depending on the source magnitude) from the best-fitting centre and set the brightest source pixels to zero. \\
    \end{itemize}
    \item \textbf{Sources with $\ks$ magnitudes between 14 and 17:}
    \begin{itemize}
        \item [{\bf Steps 1 and 2:}] Segment, deblend, and create and check the fitting mask in the same manner as in (ii). \\
        \item [{\bf Step 3:}] Fit the same PSF models in the fitting mask as in (ii). \\
        \item [{\bf Step 4:}] Remove the source using the model that minimises the least-squares error within a radius of 10 to 15 pixels (depending on the source magnitude) from the best-fitting centre and set the brightest source pixels to zero. \\
    \end{itemize}
\end{enumerate}

In most cases, the tailored Moffat or Compound model minimises the error. However, in very crowded or nebulous regions, the average Moffat or Compound models minimise the error because oversubtraction can occur owing to source and background confusion. Further, if there is some fitting error, then the average models minimise the error. 

\subsubsection{Faint point-source removal}
\label{sec:faint_removal}
Now we describe the faint-source removal steps. The process is similar to that described in the previous section, except we do not mask the brightest pixels. We fit PSF models (with parameters described in detail in \S\ref{sec:analytic_psf_models}) using the Levenberg--Marquart algorithm. As initial guesses for the algorithm, we use the median value of the faint PSF parameters from \S\ref{sec:faint_modeling} in the same area of the image. To do this, we divide the image into 56 grid cells and calculate the median parameters and standard deviation of parameters in each bin (Figure \ref{fig:faint_psf} shows how we grid the image). We use the same PSF models as in the previous section. The steps we follow are described in Figure \ref{fig:remove_stars} and are as follows: 

\begin{enumerate}[\bfseries (i)]
    \item {\bf Bright stars that do not have enough pixels in the fitting mask from the previous section:}
    \begin{itemize}
        \item [{\bf Step 1:}] The cutout and segmented/deblended cutout is carried on to this step from the previous section, depending on the magnitude of the star.\\
        \item [{\bf Step 2:}] The new fitting mask is now composed of the segmented, deblended pixels belonging to the source we found. We check if the fitting mask has at least 12 pixels. If it has fewer, we create a circular fitting mask with a radius of 7 to 8 pixels. \\
        \item [{\bf Step 3:}] Then, the four PSF models from the previous section are fitted within the fitting mask. The centre must be located within two pixels from the cutout's centre and all parameters can vary by 2$\sigma$ from the median of the local faint parameters. \\
        \item [{\bf Step 4:}] Then, we remove the source using the model that minimises the least-squares error within a radius of 10 pixels.
    \end{itemize}
    \item {\bf Faint stars with a catalogue magnitude between 17 and 22:}
    \begin{itemize}
        \item [{\bf Step 1:}] First, a cutout is made with a size indicated in Table \ref{tab:star_stats}. The cutout is segmented with groups of 5 pixels above a threshold of 2$\sigma$ (panels C1 and C2 in Figure \ref{fig:remove_stars} show an example of a cutout and segmented/deblended cutout of this magnitude range). If no segmented source is found at 2$\sigma$, no source is removed.\\
        \item [{\bf Step 2:}] The fitting mask is created from the deblended, segmented pixels (panel C3 in Figure \ref{fig:remove_stars} shows an example of a fitting mask). If the fitting mask contains fewer than 12 pixels, a circular fitting mask is created with a radius of 3 pixels. \\
        \item [{\bf Step 3:}] Then, the same four PSF models with the same bounds as in the previous case are fitted (panel C4 shows the residuals after using these four models in Figure \ref{fig:remove_stars}). \\
        \item [{\bf Step 4:}] Then, we remove the source using the model that minimises the least-squares error within a radius of 5--6 pixels (panel C5 shows the result in Figure \ref{fig:remove_stars}). 
    \end{itemize}  
\end{enumerate}


\subsection{Residual image}
\label{sec:residual_image}

\begin{figure*}
\includegraphics[width=.95\linewidth]{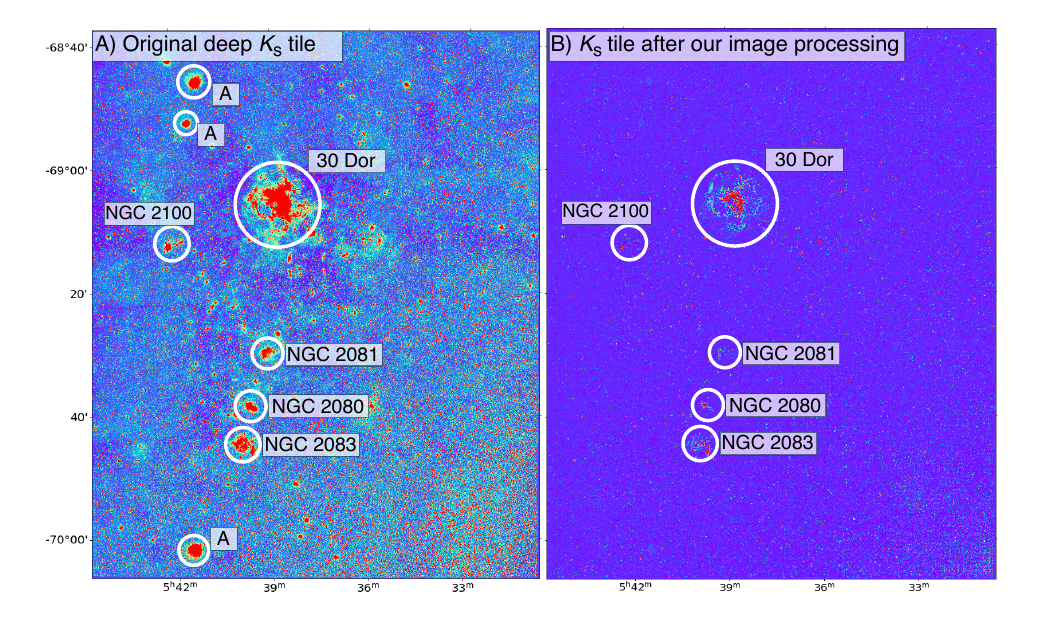}
  \caption{(\textit{Left}) Deep VMC $\ks$ tile LMC 6$\_$6 before applying our image processing. We label saturated Milky Way stars with the letter ‘A;' these are removed during our image processing. We also label five young star-forming regions. (\textit{Right}) Tile after being processed by our image pipeline (the residual image). We label the same five star-forming regions, which all have some extended component that remains in the residual image.}
  \label{fig:before_after}
\end{figure*}

After removing all point-like sources, we have a residual image. We show the original deep tile image and the corresponding residual image in Figure \ref{fig:before_after}. We show some objects we remove, including saturated Milky Way stars and image artefacts. We also show five star-forming regions. These regions all have an extended component that remains in the residual image. 

\section{Isophote detection and characterisation} 
\label{sec:isophotes}
To extract sources from the residual image, we apply isophote detection using $\textsc{scikit-measure}$ contours at a local threshold of 3$\sigma$ (based on the noise image we calculated in \S\ref{sec:star_remove}). In the next sections, we describe how we calculate their structural parameters, their integrated photometry, how we cull spurious objects, how we split up the 3$\sigma$ detections into higher-significance isophotes, and, finally, how we add expansive counterparts to the initial 3$\sigma$ detections. 

\subsection{Structural parameters}
\label{sec:structural_parameters}
Initially, there are 1,207,659 isophotes detected at 3$\sigma$. The choice of 3$\sigma$ is somewhat arbitrary. We also experimented with detecting at 5$\sigma$ and 10$\sigma$, but we found we recover most known star clusters at 3$\sigma$. However, this also means there are more spurious detections that must be removed.  We use the isophote vertices to calculate isophote geometric radii, $r_G$ (based on a circle with the same area as the isophote), and geometric centroids, ($x_G, y_G$). There are many small, unphysical isophotes at this stage; therefore, we remove the smaller objects. If $r_G$ is less than 2$\times$ the local FWHM, we remove it. The local FWHM is calculated in 56 bins across the image, as in Figure \ref{fig:faint_psf}. We also remove all isophotes closer than 75 pixels to the edge of the tile. After this step, there are 4,771 isophotes left. 

For each remaining isophote we next compute a series of light-weighted quantities. For a given quantity $s$, we define its light-weighed average over a contour by
\begin{equation}
\begin{split}
\braket{s} = \frac{\sum_{i} s  I(x_i,y_i)}{\sum_{i} I(x_i,y_i)},
\label{eqn:observable}
\end{split}
\end{equation}
where $I(x,y)$ is the intensity of the pixel at position $(x,y)$, and the summation runs over all pixels $(x_i,y_i)$ within a given contour. The light-weighted quantities we compute are the position $(\langle x_0\rangle, \langle y_0\rangle)$ (defined by $s=x$ and $s=y$, respectively), the radius $r_L = \sqrt{(x-\langle x_0\rangle)^2 + (y - \langle y_0\rangle)^2}$ and the inertia tensor $\langle \mathbf{T}\rangle$, whose entries are $T_{xx} = (\langle x \rangle - \langle x_0\rangle)^2$, $T_{xy} = T_{yx} = ( \langle x \rangle - \langle x_0 \rangle)(\langle y \rangle - \langle y_0 \rangle)$ and $T_{yy} = (\langle y \rangle - \langle y_0\rangle)^2$. The larger and smaller eigenvalues of $\langle \mathbf{T}\rangle$ define, respectively, the major and minor axis lengths, $a$ and $b$, and the orientation of the eigenvector associated with eigenvalue $a$ defines the position angle, $\theta$. Finally, $a$, $b$, and $\theta$ define an ellipse, to which we refer hereafter as the inertia tensor ellipse; we calculate the median pixel value inside this ellipse, which we will use below to cull spurious objects.

\subsection{Integrated photometry}
\label{sec:integrated_photometry}
Next, we calculate the $YJK_\text{s}$ integrated photometry of our 3$\sigma$ isophotes. To do this, we take the $Y$, $J$, and $\ks$ deep tiles and apply background filtering and estimate the background and noise in the same way as in \S\ref{sec:bkg_noise}. We point the reader to \citet{cioni2011} for further details on the $Y$ and $J$ observations. We apply the nebuliser with the (90, 30) filter sizes, because this is very similar to how CASU and the WFAU apply it when they perform photometry on the VIRCAM images. Then, for each isophote we calculate the photometry inside the isophote vertices. To do this, we select the pixels that fall into the isophote boundaries, sum them, and then convert them to a magnitude. We use the zero points and exposure times in the deep tile headers to calculate the VISTA magnitudes. The zero points of the images are recalculated every time an image is stacked or mosaicked. Then, we convert VISTA to Vega magnitudes using the transformations from \citet{gonzalez2017}. We also calculate the magnitudes in the same manner on the residual $\ks$ image; this magnitude will help us cull spurious objects in the next section. Finally, we also calculate the surface brightness and the peak surface brightness for each isophote in $YJ\ks$. Therefore, we have three integrated magnitudes from the $YJ \ks$ tiles, three surface brightness values from the $YJ \ks$ tiles, peak surface brightness values from the $YJ \ks$ tiles, and an integrated magnitude from the residual $\ks$ image. 

\subsection{Disposal rules}
\label{sec:rules}

\begin{figure*}
  \includegraphics[width=.8\linewidth]{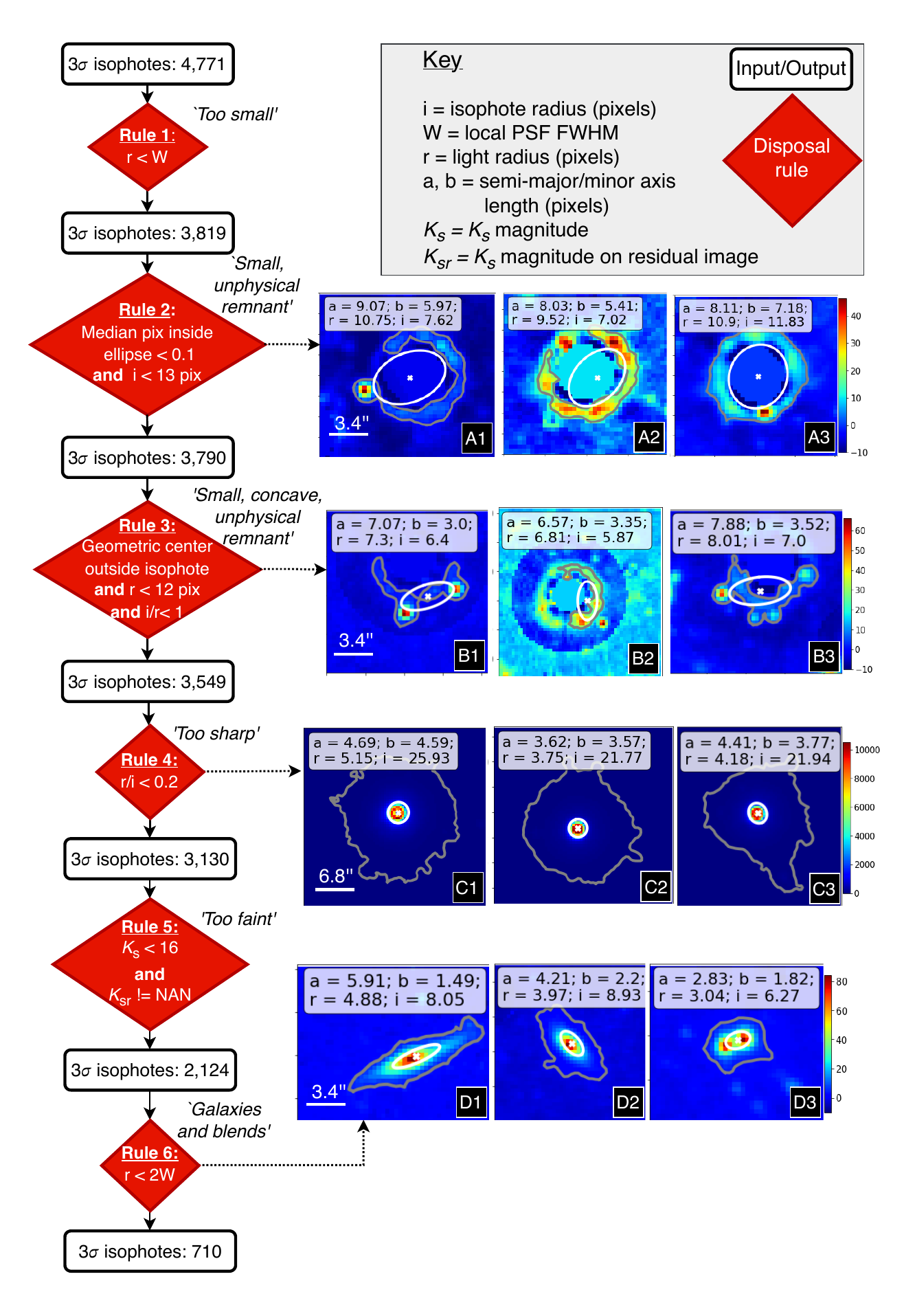}
  \caption{Diagram showing the six disposal rules for the isophotes, which displays how we cull most of the spurious isophote detections. In panels A, B, C, and D, the grey boundary is the isophote, the inertia tensor ellipse is shown in white, and the white `X' is the light-weighted centre. The panels labelled A are detections around poorly subtracted stars. These objects' inertia tensor ellipse has a median pixel value of zero inside, indicating its unphysical nature. The panels labelled B are also detections around poorly subtracted stars. These differ from those in A, because generally these are relatively bright objects that are revealed after subtraction of a bright star. We remove them because it seems that they are usually point-like sources whose size is overestimated by their mixing with leftover emission from the bright star. The panels labelled C are very bright stars that were left in the image. Finally, the panels labelled D1 and D2 appear to be extended, edge-on background galaxies, whereas plot D3 appears to be a stellar blend.  }
  \label{fig:rules}
\end{figure*}

Now we cull spurious objects by removing objects that violate one of our six disposal rules, which we outline in Figure \ref{fig:rules} and list below:

\begin{itemize}
  \item [{\bf Rule 1:}] The `too small' rule: $r_L$ $<$ local FWHM.\\
  \item [{\bf Rule 2:}] The small, unphysical remnant rule: median pixel inside inertia tensor ellipse $<$ 0.1 and $r_G$ $<$ 13 pixels. Plots A in Figure \ref{fig:rules} show three examples removed at this stage. \\
  \item [{\bf Rule 3:}] The small, concave, unphysical remnant rule: geometric centre ($x_G$ and $y_G$) outside the isophote boundaries, $r_L$ $<$ 12 pixels and $r_G / r_L$ $<$ 1. Plots B in Figure \ref{fig:rules} show three examples.\\
  \item [{\bf Rule 4:}] The `too sharp' rule: $r_L / r_G <$ 0.2. Plots C show three examples.\\
  \item [{\bf Rule 5:}] The `too faint' rule: $\ks$ magnitude $>$ 16 and the residual $\ks$ magnitude not equal to `NaN'.\\
  \item [{\bf Rule 6:}] The galaxies and blends rule: $r_L$ $<$ 2$\times$ the local FWHM.
\end{itemize}

The first rule is to remove objects smaller than the local FWHM. In general, $r_L$ is smaller than $r_G$ (except for very irregularly shaped isophotes). Therefore, we initially apply a conservative size cut on $r_L$. This size cut effectively removes many small and spurious detections. 

The second rule is based on experimentation and a study of the parameters of isophotes around poorly subtracted stars. The first step is implemented because if the inertia tensor ellipse has a median pixel value of zero or less, this indicates that the isophote is wrapped around a poorly subtracted/masked star. This is so, because when we mask pixels in the image pipeline, we set high-intensity pixels to zero. 

The third rule is chosen because if a small isophote is wrapped around a poorly removed star, the geometric centre will be outside the isophote. Because this occurs for objects we want to keep, we also specify that the ratio of the geometric radius to light radius is $<$1. This helps us remove more spurious objects, because generally the geometric radius is greater than the light radius. We note that in rule three, it is debatable whether we should remove these objects. This is because often we remove a star and then objects in the image show up that were not detected before. Based on looking at many of these objects, we decided to remove them, because we believe that the objects that are newly revealed are mixing with residual light left over from the poorly removed star, which artificially increases its size. As a result, such objects are not removed by rule one, but they should be. 

The fourth rule is designed to remove stars that remain in the residual image. When we remove stars from the image, we do so based on the stellar probability and sharpness values from the PSF catalogue. However, some probable stars that do not match our selection criteria are left in the image. We can identify these as stars based on the ratio of the light radius to the geometric radius being very small. This is because the light of stars is sharply peaked. 

The fifth rule cuts out objects that are very faint. This helps remove background galaxy contamination. 

Finally, the sixth rule is a final size cut. This size cut might remove legitimate detections which we would like to keep, but it helps to remove spurious detections like blended stars and background galaxies. 

We order the disposal rules to assess how each step affects our matching with other cluster surveys. Therefore, first we want to remove spurious detections and single stars from our list (rules one, two, and three). Then, we can assess the effect of removing faint and small objects by seeing how those cuts affect our matches with other surveys. \S\ref{sec:comparison} describes our matching with other surveys at each stage of culling. After removing spurious detections and making some size cuts, we have 710 detections at 3$\sigma$. 



\subsection{Splitting initial detections into higher-significance isophotes}
\label{sec:higher_sigma}

\begin{figure}
\centering 
  \includegraphics[width=.85\linewidth]{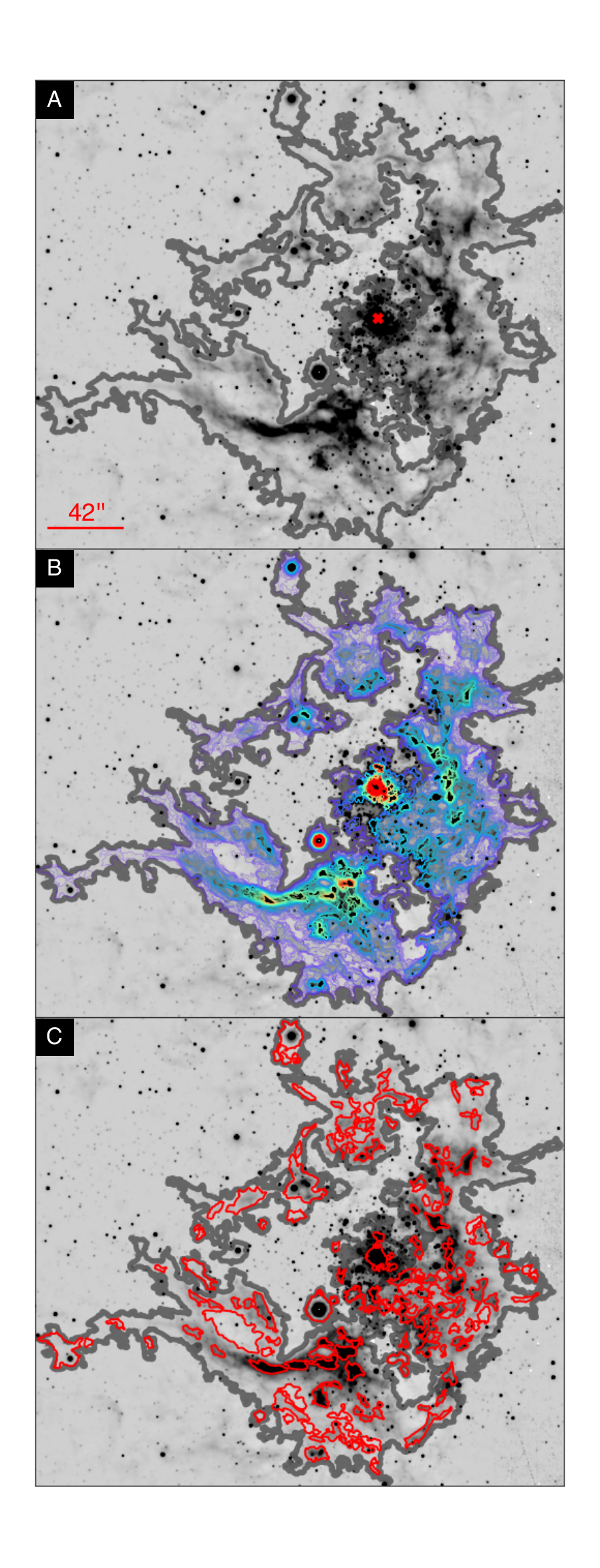}
  \caption{Example of the largest multi-star structure we detect, the 30 Doradus star-forming complex. 42 arcsec is equivalent to 10 pc at the LMC's distance. This figure shows that we can detect large star-forming complexes, study their complex structure, and find nested star clusters. Panel A shows a grey-scale image cutout around the isophote that coincides with 30 Doradus. The red `X' indicates the location of the star cluster R136 at the core of 30 Doradus. Panel B shows the higher-significance isophotes starting at 10$\sigma$ and ascending; there are 2,241 contours. They are shown as rainbow contour lines. Panel C shows the 156 atomic isophotes in red. The atomic isophotes are described in \S\ref{sec:higher_sigma} and Figure \ref{fig:isophote_terminology}.}
  \label{fig:30_dor}
\end{figure}


\begin{figure}
  \includegraphics[width=.99\linewidth]{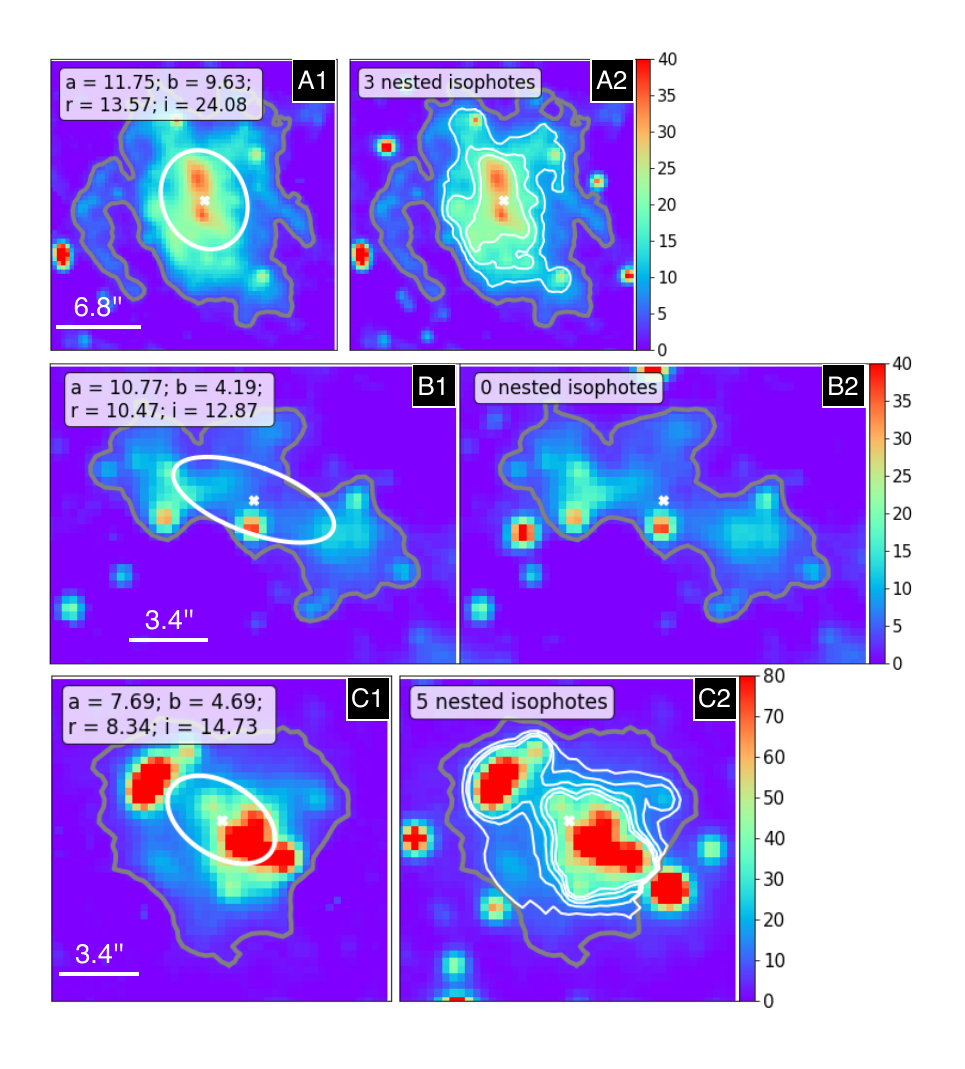}
  \caption{Three examples of the initial 3$\sigma$ isophotes (left-hand panels) and their higher-significance, nested isophotes (right-hand panels). The left-hand panels are based on the residual $\ks$ image and the right-hand panels are from the image with only Milky Way stars removed. The grey line is the isophote boundary, the white line is the inertia tensor ellipse and the white cross is the light-weighted centre. The calculation of the inertia tensor ellipse is described in \S\ref{sec:structural_parameters}. Also labelled are the semi-major and semi-minor axis lengths, $a$ and $b$, the light-weighted radius, $r$, and the isophote geometric radius, $i$. In the right-hand panels, the higher-significance isophotes are shown in white. All axes are in units of pixels and colour bars are in units of background-subtracted counts. Example B has no higher-significance isophotes, whereas examples A and C have one or more, which shows the breadth of objects we can detect and characterise.}
  \label{fig:higher_isophotes}
\end{figure}

The 710 3$\sigma$ isophotes serve as our initial detections. In order to study the hierarchical properties of the isophotes, as well as split the ones with multiple peaks into multiple detections, we split all isophotes into smaller, higher-significance isophotes. To do this, first we use a residual image with only Milky Way foreground stars removed. Next, we look at each 3$\sigma$ isophote in the catalogue and consider higher-significance isophotes inside this 3$\sigma$ isophote, from 10$\sigma$ and ascending in increments of 5$\sigma$. 

After detecting the higher-significance isophotes, we calculate the same structural parameters as in \S\ref{sec:structural_parameters}. We only keep those with $r_G$ $>$ 2$\times$ the local FWHM. Next we calculate the $YJ \ks$ photometry in the same way as in \S\ref{sec:integrated_photometry}. Then, we take the higher-significance isophotes resulting from disposal rules 1 and 4. This removes small, spurious detections and stars. We do not apply rules 2 and 3 because we do not need to: we are using the residual image with only the Milky Way stars removed at this stage. We do not apply rules 5 and 6 either. Those rules are designed in part to remove small background galaxies, which have a lower probability of remaining at this stage. After applying the disposal rules, there are 10,005 higher-significance isophotes inside 390 of the 710 $3\sigma$ isophotes; the remaining 3$\sigma$ isophotes do not contain any higher-significance detections. 

For every isophote tree, we determine its `atomic isophotes.' The atomic isophotes are the isophotes that do not separate into more than one isophote at each higher isophote level; therefore, the atomic isophotes can have children, but they are all on the same branch. The atomic isophotes are important for separating the isophotes into discrete groups that are likely nested star clusters. There are 1,055 atomic isophotes. 

\begin{figure}
\centering
  \includegraphics[width=.95\linewidth]{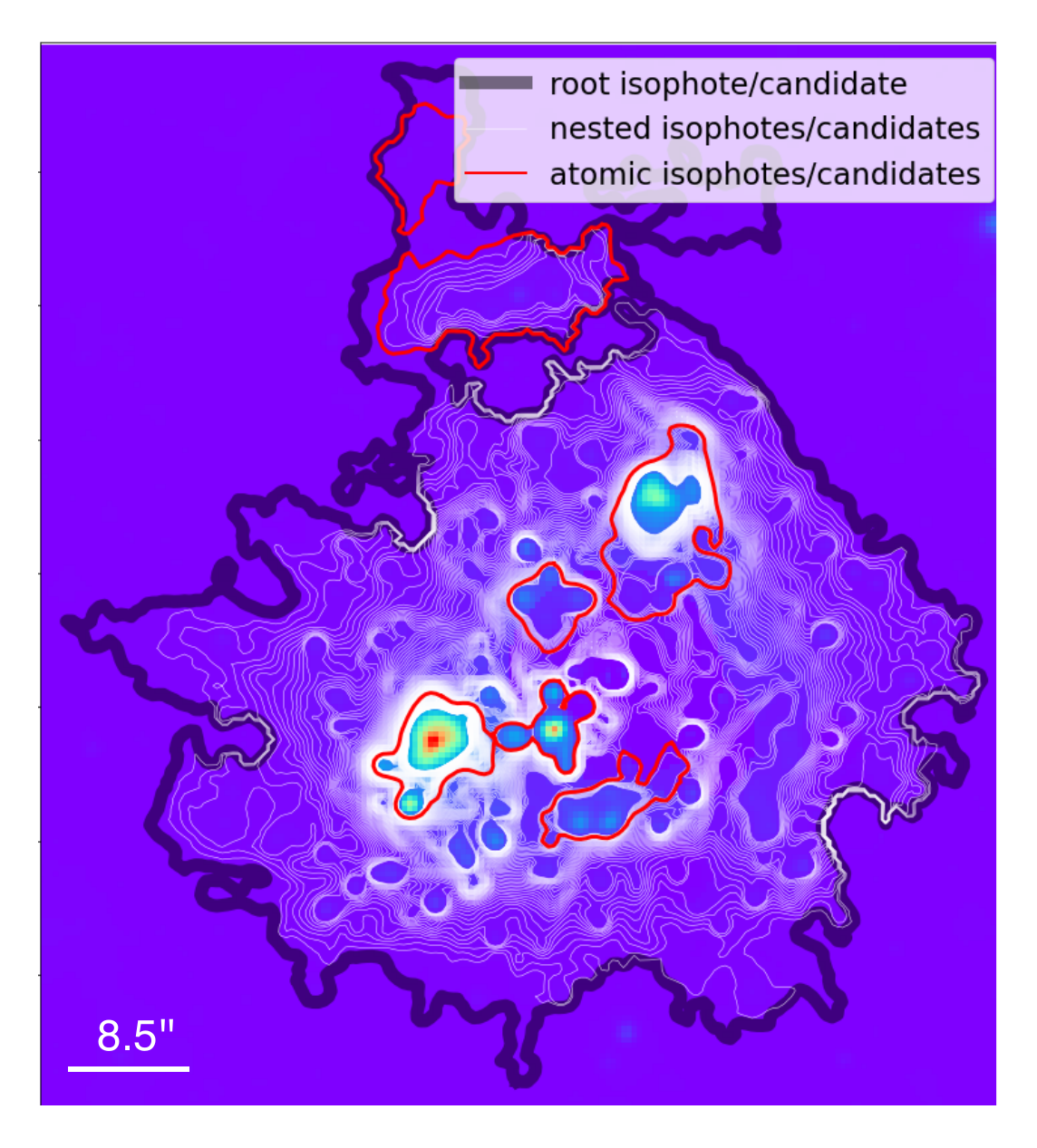}
  \caption{This Figure shows the isophote/cluster candidate terminology we adopt. The black isophote is the initial 3$\sigma$ isophote detection. We refer to these as root isophotes or candidates. The higher-significance isophotes are shown as white. We refer to these as nested isophotes or candidates. Finally, the atomic isophotes are shown in red; we refer to these as atomic isophotes or candidates.}
  \label{fig:isophote_terminology}
\end{figure}

From henceforth, we will refer to the initial detection at 3$\sigma$ as the `root' isophotes/candidates, the higher-significance isophotes as higher-significance or nested isophotes/candidates, and the atomic isophotes as atomic isophotes or candidates. To clarify this terminology, we show Figure \ref{fig:isophote_terminology}, which shows an example of how we define the isophote tree and each isophote level.


We show the largest isophote, which coincides with the 30 Doradus star-forming complex, in plot A of Figure \ref{fig:30_dor}. In plot B of Figure \ref{fig:30_dor}, 
we detect its 2,241 higher-significance isophotes. In plot C of Figure \ref{fig:30_dor}, we detect its 156 atomic isophotes. Figure \ref{fig:higher_isophotes} shows four examples of the initial 3$\sigma$ detections and their higher-significance nested isophotes. Figures \ref{fig:30_dor} and \ref{fig:higher_isophotes} show the wide ranges of morphology and size of objects that our method is able to detect. 


\subsection{Adding expansive counterparts}
\label{sec:expansive}
\begin{figure*}
  \includegraphics[width=.9\linewidth]{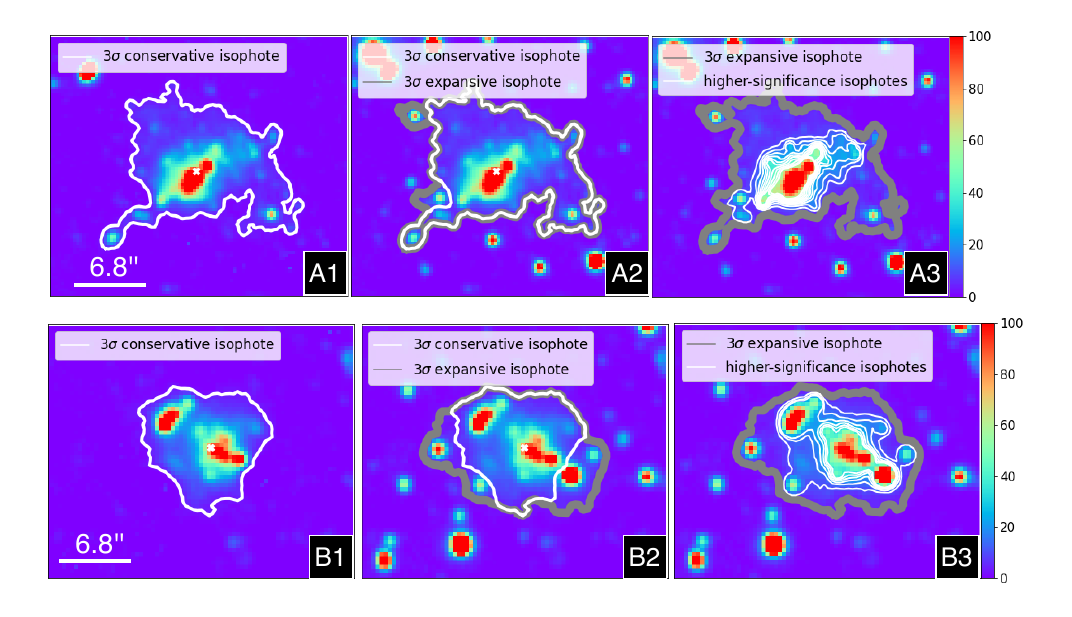}
  \caption{Examples showing our procedure for producing the expansive candidates, described in \S\ref{sec:expansive}. Each row is a different example and each of the middle panels show how we add in neighboring point-sources to the conservative detections. Reading down the left-most column, A1 and B1 show the residual $\ks$ image with the 3$\sigma$ conservative isophote in white. Reading down the middle column, A2 and B2 show the 3$\sigma$ conservative isophote in white and the 3$\sigma$ expansive isophote in grey. A3 and B3 show the 3$\sigma$ expansive isophote in grey and its higher-significance isophotes in white. All axes are in units of pixels and all colour bars are in units of  background-subtracted counts. }
  \label{fig:expansize}
\end{figure*}

Our detection algorithm removes point sources from the image and detects those objects that are extended. Therefore, we remove point sources that are part of star clusters and star-forming regions. Given the data available, there is no obvious way to ascertain whether the neighbouring point sources are single bright stars that are physically associated with but lie at the edge of the detected multi-stellar structure, or if they are simply chance alignments. 

To deal with this ambiguity, we add `expansive' counterparts to the initial `conservative' detections. To do so, we first take the residual $K_\text{s}$ image with only the Milky Way stars removed, as opposed to the full residual image with both Milky Way and LMC stars removed. We then detect isophotes at 3$\sigma$ in this image. We retain only those isophotes containing at least one of our 3$\sigma$ `conservative' isophotes. We follow this same procedure to obtain the higher-significance detections, starting at 10$\sigma$ and going upwards. We finally compute the structural parameters and $YJK_\text{s}$ photometry for all detections. Two examples of conservative isophotes and their expansive counterparts are shown in Figure \ref{fig:expansize}.

After detecting the expansive isophotes, we use the 3$\sigma$ expansive isophote parameters to remove remaining detections around poorly subtracted/masked stars. Disposal rules one and two did not remove them all. Now, with the expansive catalogue, we can remove them, because the expansive catalogue isophote encompasses the bright star that the conservative isophote is near. Therefore, we remove expansive objects using rule four. After this set of removals, we are left with 682 conservative isophotes; these 682 conservative isophotes contain 9,955  higher-significance isophotes (inside 362 of the conservative isophotes), and 1,010 atomic isophotes. There are 485 expansive isophotes and 15,422 of their higher-significance isophotes. We also detect the atomic isophotes of the expansive isophotes; there are 1,114 of them. There are fewer expansive isophotes because they are larger and sometimes encompasses many conservative isophote detections.

\subsection{Construction of the conservative and expansive star cluster candidate samples}
\label{sec:catalogues}
The 682 3$\sigma$ conservative isophotes and their 9,955 higher-significance isophotes form our conservative sample of star cluster candidates. There are 1,010 atomic isophotes in the conservative sample. 

The corresponding expansive sample is composed of the expansive isophotes and the 47 conservative isophotes that do not have an expansive counterpart. Combining these, there are 537 3$\sigma$ detections, 15,478 nested higher-significance isophotes, and 1,161 atomic isophotes that form our expansive sample of star cluster candidates.

\subsection{Descriptive statistics}
\label{sec:discussion}

\begin{figure*}
  \includegraphics[width=.9\linewidth]{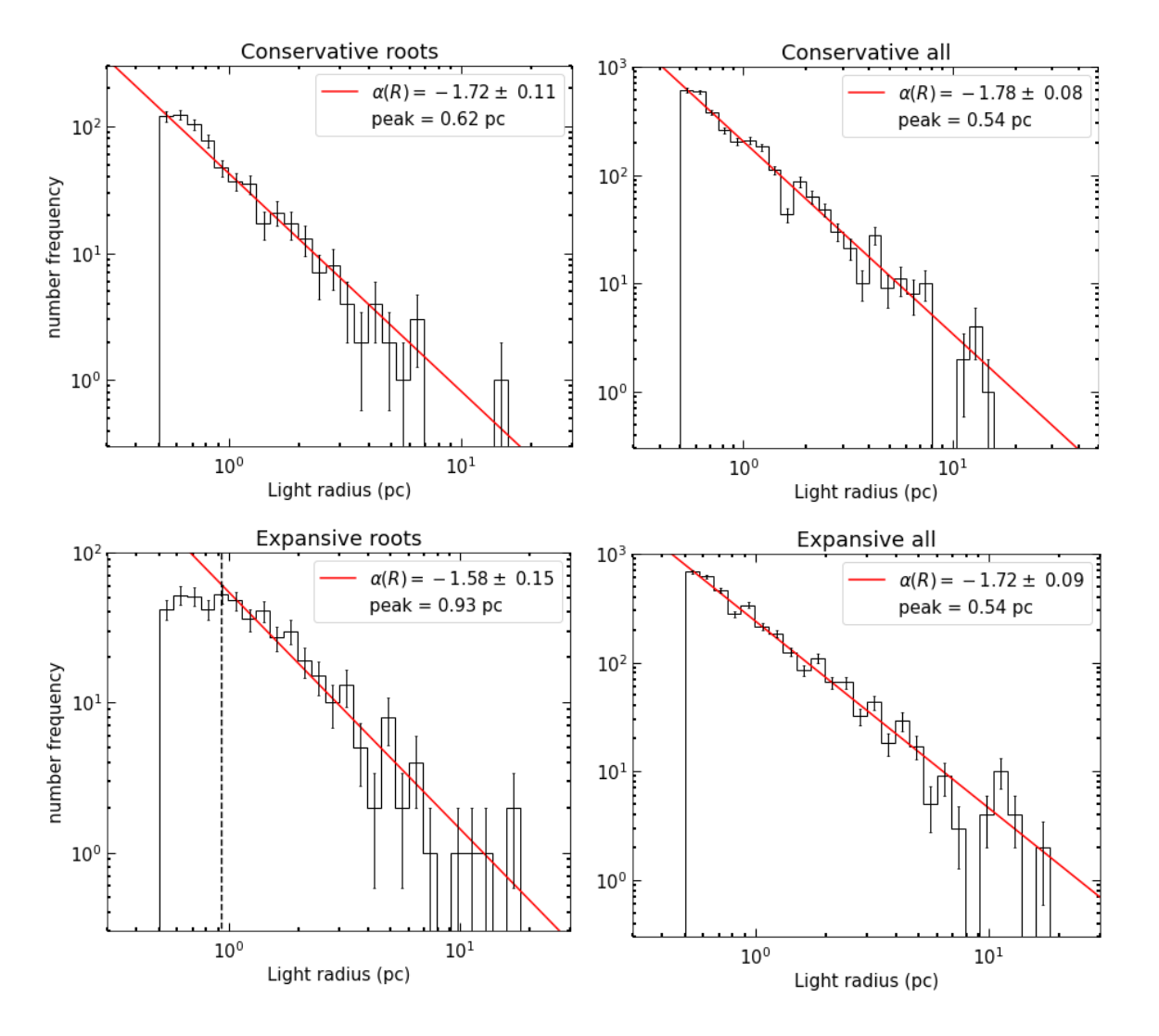}
  \caption{Here we show the distribution of the light-weighted radii ($r_L$). At top left are the 3$\sigma$ roots of the conservative sample, which are the base detections of our sample. At the top right are all isophotes from 3$\sigma$ and above, at bottom left are the 3$\sigma$ roots of the expansive sample, and at bottom right are all expansive isophotes from 3$\sigma$ and above. Also indicated in the plots are the peaks of the distributions and the power-law fits and their errors, plotted from the peak of the distribution to the tail. The error bars represent Poissonian error. The power-laws all follow a slope of less than two, which is indicative of a scale-free, fractal distribution. The power-law slopes of the conservative groups and the expansive groups agree within 1$\sigma$.}
  \label{fig:size_dist}
\end{figure*}

Here, we display several representative descriptive statistics of the conservative and expansive samples. In Figure \ref{fig:size_dist}, we plot size distributions of the conservative roots (the 3$\sigma$ candidates), all conservative candidates (3$\sigma$ and higher), the expansive roots (the 3$\sigma$ candidates), and all expansive candidates (3$\sigma$ and higher). We fit power laws to them, from the peak to the tail. All follow power-law exponents 
of less than 2, indicative of scale-free, self-similar distributions. All peaks of the size distributions are about 2$\times$ larger than the width of the PSF, which is the size cut we implemented. 

\section{Discussion} 
\label{sec:discussion}

In this section, we first discuss contaminants present in our star cluster catalogue and estimate their contamination rates. Next, we do a thorough comparison with star cluster catalogues in the literature. Finally, we compare our detections with publicly available $\textit{JWST}$ imaging data.

\subsection{Contaminants}
\label{sec:contaminants}

\subsubsection{Background galaxies}
\label{sec:galaxies}
To estimate the contamination rate by background galaxies, we look at galaxy counts from the Great Observatories Origins Deep Survey (GOODS; \citealt{goods2010}). The survey covers areas of 172.5, 159.6 and 173.1 square arcmin, respectively, the $J$, $H$, and $K_\text{s}$ bands, . The authors performed photometry and estimated galaxy counts in bins of 0.25 magnitudes in all of their near-infrared bands. We converted their AB magnitudes to the Vega system using the transformations provided. Based on their reported cumulative galaxy counts, we expect approximately 374 galaxies in the 1.77 square degree region of sky covered by our sample and in the same magnitude range as our detections. This is an upper limit, because an unknown number of galaxies will fall below our size cuts given that we only keep objects up to 2$\times$ the PSF width. 

In order to more accurately calculate the contamination rate, we use a galaxy catalogue from \citet{bell2022}. They selected likely background galaxies across the LMC from VMC PSF sources with $K_\text{s}$ $\tt{SHARP}$  $>$ 0.5 (i.e., with the sharpness indicating that the object is unlikely a single star) and $(J-K_\text{s}) > 0.8$ mag. Then, they performed aperture- and PSF-matched photometry from the optical to the near-infrared, constructed spectral energy distributions and fitted them to 62 galaxy templates and 200 stellar templates; they classify objects for which the minimum $\chi^2$ value corresponds to a galaxy template as more likely to be galaxies than stars or blended stars. However, in our tile LMC 6$\_$6, owing to crowding and nebular emission, many objects that have galaxy-like parameters in their catalogue are actually star clusters or parts of star-forming regions. Therefore, we cannot simply cross-match with their catalogue and remove the matches. Instead, we adopt a more conservative approach. We downloaded their table and found sources with minimum $\chi^2$ values corresponding to galaxy templates, cross-matched them with the VMC PSF catalogue, and only kept objects with $K_\text{s} < 16.5$ mag, $\tt{Star\_prob}$ $<$ 0.33, and absolute $K_\text{s}$ error $< 0.15$ mag. The magnitude threshold was chosen to be conservative, so we go half a magnitude fainter than our star cluster candidates. Then, the stellar probability value is chosen to be low, indicating the source is likely a blend or a galaxy. Finally, we choose to make a cut on magnitude error, because we found too many galaxies in the field that seemed spurious without this cut. There are 749 objects meeting these criteria in the tile region. Based on the GOODS galaxy counts, it is unlikely that these objects are all galaxies, but we have also selected galaxies half a magnitude deeper in $\ks$, to be on the safe side. 

We next check how many of these 749 candidate galaxies are within the boundaries of our isophotes that define candidate star clusters. There are 70 matches with our conservative sample, containing 682 isophotes. This implies a contamination rate of 10\% in our sample. However, looking at the matches, many galaxy candidates overlap with star clusters or star-forming regions. Therefore, we do not remove the matches from our sample. We use this information to suggest that there is an upper limit to the contamination rate of 10\% in our sample. 

\subsubsection{Blended sources}
\label{sec:blended}
The second major source of contamination in our sample is blended sources left in the image after point-source removal. These sources are composed of two or more sources that are not temporally or spatially related in 3D space, but which are close enough in angular projection that the telescope PSF causes them to appear as a single, extended source. As discussed by \citet{bell2019,bell2022}, such sources are also significant contaminants in galaxy catalogues, but they can partly be separated from galaxies based on colour, since blended stars generally have bluer colours while galaxies are redder.

To estimate the effect of blended sources on our sample, first we find the density of PSF stellar-like sources in the tile in our isophote magnitude range. There are 62,242 sources between $K_\text{s}$ magnitudes of 8 and 16.5 with stellar-like PSF paramters, yielding a density, $D$, of objects in the tile of $5.353 \times 10^{-3}$ arcsec$^{-2}$. Next, we must estimate how close two point sources must be for them to appear as a blend in our analysis pipeline. To make this estimate, we assume that the intrinsic distribution of stars is Poissonian, so that, in the absence of blending, the distribution of nearest-neighbour angular distances for all stars would follow a Poissonian distribution; the effect of blending is to cause a deviation below such a distribution at small separations, where what should be multiple point sources instead are extracted as a single extended source. Thus we can infer the separation at which blending becomes significant by measuring the scale below which the separation distribution becomes sub-Poissonian. We find that this distance is one pixel, or approximately 0.34 arcsec. We use this distance as the radius to compute a circular area, $A_\text{blend}$, such that if more than one source is located inside an area of $A_\text{blend}$ then it will cause a blend in our sample. The probability that any given point source will be blended, again assuming a Poissonian distribution of stellar positions, is therefore $P_\text{blend} = 1 - \exp({A_\text{blend} D}) = 9.854 \times 10^{-4}$, and the density of blends on the sky is $D_\text{blend} = P_\text{blend} D = 2.674 \times 10^{-6}$ arcsec$^{-2}$. The density of isophotes in the tile is $D_\text{isophotes} = 9.26 \times 10^{-5}$ arcsec$^{-2}$, suggesting that the blend contamination fraction $ D_\text{blend} / D_\text{isophotes} = 0.0289$, or approximately 3\%. 

One could argue that our calculated blended contamination fraction is an overestimate or underestimate of the true contamination fraction of the tile image considered. An argument for it be an underestimation in some regions of the tile image would be that LMC tile $6\_6$ has a varying stellar density, with much higher density around star forming regions such as 30 Doradus. However, our definition of stellar blends only includes blends of unrelated sources; therefore, our calculation does not include blends of related sources. An argument for it being an overestimation of the true contamination fraction is that since we consider stars over a magnitude range of 8.5 magnitudes, a blend of one of the brighter with one of the fainter of these would probably go unnoticed. So the actual contamination would be less severe. 

\subsection{Comparison to previous studies}
\label{sec:comparison}

We next compare our automatic sample to two star cluster catalogues that were composed using primarily by-eye methods and one of higher-resolution-confirmed star clusters. Therefore, we show we can detect star clusters that have been found by eye and also confirmed star clusters. 

\subsubsection{Comparison to Romita et al. (2016)}
\label{sec:romita}

\begin{table*}
	
	\label{tab:parameters}
	\begin{tblr}{XXXXX}
		\hline
             & conservative:  & conservative: & conservative: &  expansive
		  \\
            & after rule 4 & after rule 5 & after rule 6 &  
		  \\
		\hline
		Number of detections & 3130 & 2124 & 682 & 537	\\
            \hline 
		\citet{romita2016}: embedded cluster candidates (67) & 62 & 62 & 55 & 56\\
            \hline
            \citet{bica2008}: candidate clusters associated with nebula (16)  &  
             15 & 15 & 13 & 14\\
		\hline 
	 \end{tblr}
\caption{Here we show our the star clusters we recover in two surveys that found clusters by-eye. The information here shows that we have more matches before employing disposal rules 5 and 6, described in \S\ref{sec:rules}.}
\end{table*}

\begin{figure}
\centering
  \includegraphics[width=.99\linewidth]{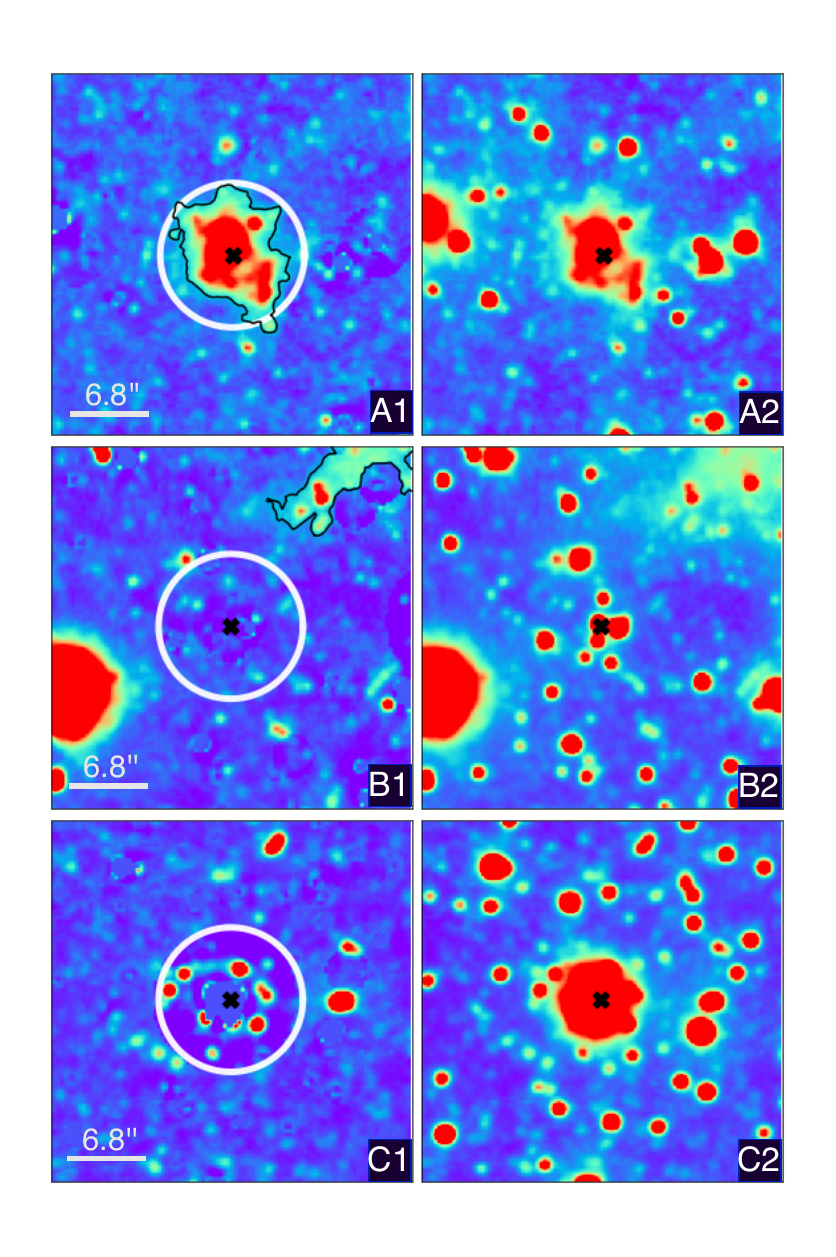}
  \caption{This figure shows three candidates from \citet{romita2016}. The left-hand images labelled A1, B1, and C1 are cutouts from the residual image. On each left-hand plot, the grey circle has a radius of 1.5 pc. The black contours are our 3$\sigma$ isophotes that form the roots of our conservative star cluster candidate sample. The right-hand images are the same, but for the raw $\ks$ image, which Romita et al. used to find the clusters by eye. On each panel, the black cross represents the centre of the cluster found by Romita et al. Panels labelled A represent a cluster from Romita et al. which we also found. Panels labelled B represent a cluster Romita et al. which we did not find, because it appears to be made up of point sources that we remove during our image processing. Panels labelled C represent a cluster from Romita et al. which we did not detect. This particular cluster is not detected because during our star removal, the sources at the centre were too close together, and our method interpreted them as one bright source, so it fitted a PSF model to many sources and masked the brightest pixels. Then, the leftover extended objects were not large enough to pass our size cut.}
  \label{fig:kr_clusters2}
\end{figure}

The first study we compare to is \citet{romita2016}, who carried out a by-eye search of the same VMC tile that we use here, focusing on regions of high CO emission. They looked for extended objects or over-densities of point-sources that overlapped with CO emission, and then classified them as embedded cluster candidates. To check the overlap between their candidates and our cluster candiates, we look at the vertices they provide to describe the footprint of their candidates; if their vertices, constructed by making a circle with their catalogue radius, intersect with an isophote in our catalogue, we consider it detected. By this criterion we detect 82\% of \citeauthor{romita2016}'s candidates in the conservative sample and 85\% in the expansive sample. Before applying disposal rule six, which is our safety measure to discriminate against contamination by blends and background galaxies, we detect 93\% of them. In Table \ref{tab:parameters}, we show that we lose eight matches of the \citet{romita2016} catalogue which we initially detected after applying disposal rule six. 

In Figure \ref{fig:kr_clusters2}, we show an example of a cluster we detect and two which we do not from \citet{romita2016}. Panel A shows a cluster we do detect. Panel B shows one we do not detect because it is comprised of point sources which our method removes. Panel C represents a cluster we did not find because during star removal, the sources were located too close together, and our method interpreted them as one source, so it fitted a PSF model to many sources and masked the brightest pixels. Therefore, the candidate shown in the panel labelled B is not an object our algorithm is designed to find, whereas the panel C candidate is more likely an object we would want to detect. However, neither of these objects are confirmed star clusters. Panel C shows a downside of the segmentation algorithm we use: it does not separate sources as well as an algorithm that uses the PSF, such as DAOPHOT \citep{Stetson1987}.

\subsubsection{Comparison to Bica et al. (2008)}
\label{sec:bica}

The next study to which we compare is \citet{bica2008}, a catalogue of likely young star clusters, associations, and nebulae, most of which were found by eye, but they also compiled the results of a number of earlier studies as well. Their catalogue is a good catalogue to which to compare because it is very comprehensive. We compare to two of their cluster types: type `CN' or `cluster in nebula' and type `NC' or `nebula with probable embedded cluster'. First, we find all of their clusters that fall within our deep tile's footprint. Then, we look at each CN and NC cluster in the deep $\ks$ image and only keep those that have sources with emission in the $\ks$ image. In the $\ks$ image, some of \citet{bica2008}'s clusters do not appear to have any matching object or emission, so we remove them. After this, there are 16 CN and 27 NC. 

Matching with our cluster samples, we detect 81\% of the CN in the conservative and 88\% of them in the expansive sample. Before applying rule six, we detect 94\% of them. We show that we lose two matches with type-CN objects after applying rule six in table \ref{tab:parameters}. We detect 100\% of the type NC in both samples. 

\subsubsection{Comparison to Milone et al. (2023)}
\label{sec:milone}

Finally, we checked the catalogue of \citet{milone2023}. They study {\sl Hubble Space Telescope} data of young star clusters. This is a good catalogue to match to because it is a catalogue of higher-resolution confirmed star clusters. There are two of their clusters in LMC tile $6\_6$. We detect both of them in both our conservative and expansive samples. 

\subsection{Comparison to available {\sl James Webb Space Telescope} images}
\label{sec:jwst}
\begin{figure}
  \includegraphics[width=.9\linewidth]{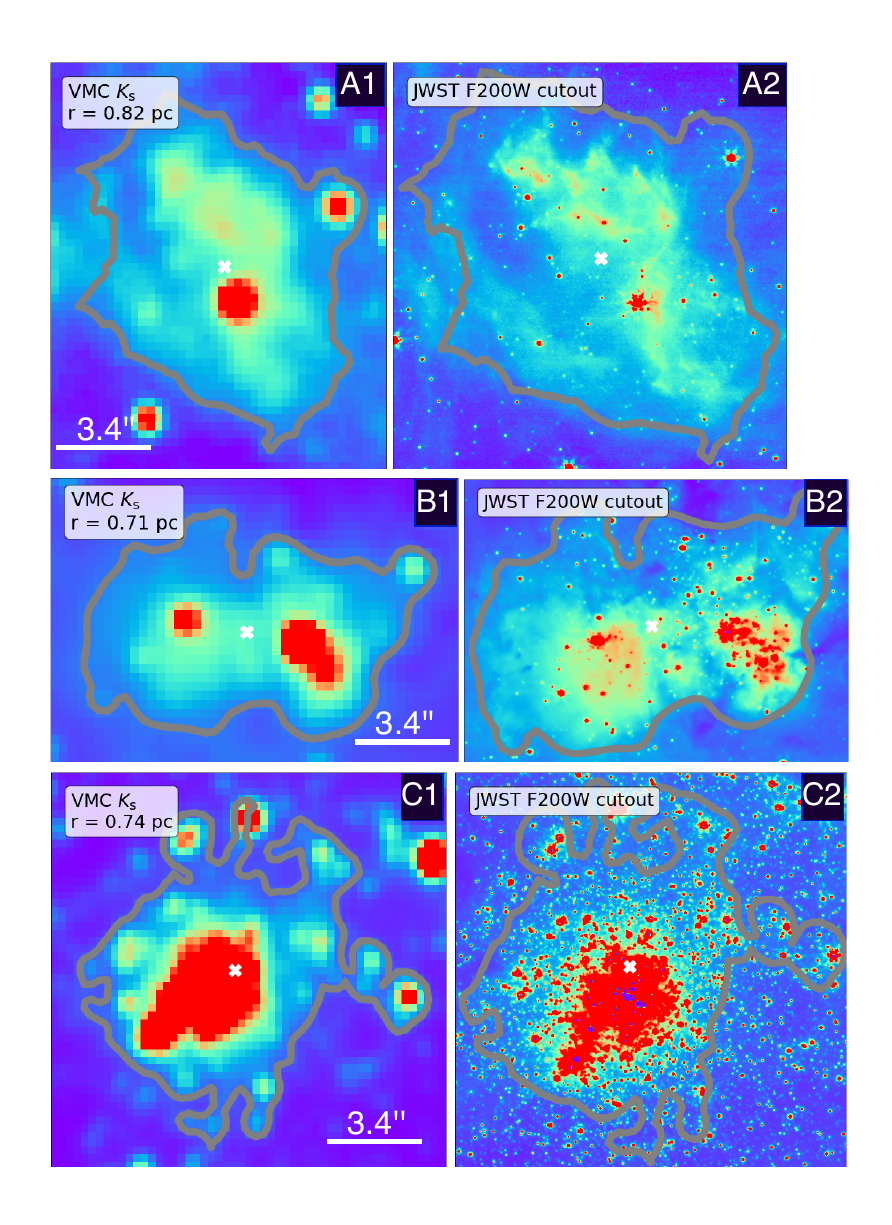}
  \caption{Three detections in the VMC $\ks$ image (A1, B1, and C1) and their corresponding images in the NIRCam F200W filter (A2, B2, and C2). The grey boundary is the isophote boundary and the white cross is the light-weighted centre. Indicated on the left plots are the light-weighted radius ($r$) converted to parsecs. These detections are examples of ones we classify as category 1. The plots labelled C correspond to the star cluster R136.}
  \label{fig:JWST}
\end{figure}

In this section, we search for our detections in publicly available {\sl JWST} data. We use publicly available {\sl JWST} NIRCam data from Early Release Observations programme 02729, which we downloaded from the MAST archive.\footnote{\url{https://mast.stsci.edu/portal/Mashup/Clients/Mast/Portal.html}} The data are described by \citet{Fahrion2023}. We use the image that corresponds to the NIRCam F200W filter, closest to $\ks$. We note that this {\sl JWST} data are targeting the 30 Doradus star-forming complex. Figure \ref{fig:30_dor} shows our isophote detection that corresponds to 30 Doradus.

We check how many of our root conservative candidates and atomic conservative candidates are completely inside the {\sl JWST} image footprint: there are 60 of the root candidates and 234 of the atomic root candidates. We note that most of the atomic candidates are inside the 30 Doradus star-forming complex (panel C of Figure \ref{fig:30_dor} shows the atomic isophotes of the 30 Doradus complex). In Figure \ref{fig:JWST}, we compare four of our cluster candidates in the VMC $\ks$ image to the NIRCam F200W.

Next, we look at the matches in the {\sl JWST} image and classify them by eye in three categories: (1) an obvious star cluster/complex; (2) a group of at least 15 point sources but whether it is a star cluster is not obvious; and (3) one to ten point sources that are clearly not a star cluster. We show the results in Table \ref{tab:jwst_matches}. 

\begin{table}
    \centering
    \begin{tabular}{llll}
        \hline
        \textbf{Cluster  } & Category 1 & Category 2 & Category 3 \\
        \textbf{candidate } &  & &  \\
        \textbf{type} & & &\\
        \hline
        root isophote  & 45 & 8 & 4\\
         conservative (60) &  &  & \\
         \hline
        atomic isophote  & 190 & 37 & 7\\
        conservative (234) &  &  & \\
        \hline
    \end{tabular}
    \caption{Table showing the type of cluster candidate we check for in the {\sl JWST} image and how many we classify in each category. Category 1 is an obvious star cluster or star forming complex; category 2 is a group of at least 15 point sources; and category 3 is one to ten point sources and clearly not a cluster. }
    \label{tab:jwst_matches}
\end{table}

These results argue that at least 80\% of our detections are genuine star clusters. They also imply that our calculation of the contamination from stellar blends is accurate.  It is worth cautioning that we cannot be certain that these detections are genuine star clusters without conducting a full search for overdensities of temporally and spatially related stars in the {\sl JWST} images; we leave this for future work. 

\section{Conclusion} 
\label{sec:conclusion}
We present the first fully-automatic detection method for semi-resolved star clusters, defined as those where the resolution is not so high that we can separate the cluster into individual stars, but is sufficient to resolve the radius of the cluster by at least a few beams. Our method is based on the idea of separating clustes from stars by filtering images to remove structures consistent with the observational PSF, and then characterising the structures that remain using structure-agnostic hierarchically-nested isophotes. The basic steps in our algorithm are as follows:

\begin{enumerate}[\bfseries (i)]
  \item We introduce a pipeline to fit the observational PSF across the field, then use this fit to remove stars.
  \item We then define isophote contours on the residual image, and using these we identify extended objects and compute their structural parameters and integrated photometry. These calculations not only allow us to characterise structures, they aid in filtering out spurious objects such as blends and background galaxies.
  \item Once we have removed interlopers, we categorise the isophotes that remain into 1) root isophotes that are the initial detections at the lowest isohpote level (3$\sigma$ above the background noise), 2) nested isophotes that are detected at 10$\sigma$ above the noise and higher in steps of 5$\sigma$, and 3) atomic isophotes that do not separate into more than one isophote at each higher isophote level. We suggest the atomic isophotes are probable nested star clusters. 
\end{enumerate}

We demonstrate this new algorithm on a tile taken from the VMC survey of the LMC, tile 6$\_$6, which contains the most actively star-forming part of the galaxy, using $K_\text{s}$ as our reference band for cluster finding. The algorithm allows us to create a new, fully automated sample of LMC star clusters. We create two such cluster samples, a conservative one with higher purity but lower completeness and an expansive one with higher completeness but lower purity, which differ in how aggressively we remove potential interlopers. The former consists of 682 root cluster candidates with 9,955 nested, higher-significance isophotes and 1,010 atomic isophotes. The latter contains 537 root isophotes, 15,478 nested isophotes, and 1,161 atomic isophotes. Isophotal sizes are power-law distributed, suggesting that we are detecting the self-similar, scale-free clustering of young stars that exists before they relax to bound clusters or disperse into the field.

We also carry out several tests of catalogue purity and completeness to validate our new method. Even for the expansive catalogue we show that maximum contamination rate should be 10\% from background galaxies and 3\% from blended stars. Comparison with the by-eye detection of 67 embedded star cluster candidates in the same VMC tile by \citet{romita2016} reveals that our method successfully detects 85\% of their candidates. Similarly, comparison with the objects classified by-eye by \citet{bica2008} as ``cluster in nebula'' and ``nebula with probable embedded cluster'' shows that our automated method recovers 88$\%$ of the former and 100$\%$ of the latter. We also detect 100$\%$ of clusters in the same region from the by-eye star cluster catalogue of \citet{milone2023}. Finally, in parts of our field for which there are publicly-available \textit{JWST} images, which are capable of resolving much finer details than our ground-based data, by-eye comparisons indicate that at least 80\% of our detections represent genuine star clusters or complexes. Taken together, these tests suggest that our automated method is capable of generating large cluster catalogues from semi-resolved images that are as good as traditional by-eye techniques, but that have the advantage of being both scalable and replicable.

While this represents a significant advance, we also identify two areas where there is room for future improvement. First, we encountered issues when removing point sources from the deep $\ks$ image, primarily stemming from the segmentation algorithm used for their separation. If we could improve this part of the method, this would enhance the completeness of our star cluster detections. Second, our method overlooks larger, looser associations and complexes within star-forming regions where the stars are loosely-packed enough that they move from the semi-resolved to the fully-resolved regime. To address this limitation, we recommend supplementing our method with a catalogue based on clustering of point sources such as that of \citet{miller2022}, allowing recovery of both semi-resolved and fully-resolved structures.

Our star cluster sample represents a unique resource with far-reaching implications. It will enable dissection of complex star-forming regions into multi-peaked detections, offering unprecedented insight into their morphological properties. The delineation given by isophote boundaries enables detailed studies of morphological shapes within these regions. Looking ahead, our methodology holds promise for broader applications. It can be readily applied to the entire VMC dataset, facilitating the detection of semi-resolved star clusters across the entire LMC. Moreover, our approach can be adapted to construct background galaxy catalogues. Lastly, our cluster sample has the potential to guide future space-based observations of young, embedded star clusters in the LMC.

\section*{Acknowledgements}

The author was supported by the International Cotutelle Macquarie University Research Excellence Scholarship. This research was supported in part by the Australian Research Council Centre of Excellence for All Sky Astrophysics in 3 Dimensions (ASTRO 3D), through project number CE170100013. MRK acknowledges funding from the Australian Research Council through Laureate Fellowship FL220100020. We thank the Cambridge Astronomy Survey Unit (CASU) and the Wide Field Astronomy Unit (WFAU) in Edinburgh for providing the necessary data products under the support of the Science and Technology Facilities Council (STFC) in the U.K. This study is based on observations made with VISTA at the ESO/La Silla Paranal Observatory under programme ID 179.B-2003. Finally, this project has made extensive use of the \texttt{python} packages: \textsc{astropy} \citep{Astropy18}, \textsc{photutils} \citep{photutils}, \textsc{shapely} \citep{shapely2007}, \textsc{scikit-learn}, \citep{scikit-learn} \textsc{matplotlib} \citep{matplotlib}, \textsc{NumPy} \citep{numpy}, \textsc{pandas} \citep{pandas} and \textsc{SciPy} \citep{scipy}.

\section*{Data Availability}
The raw data this study uses are available from the ESO Science
Archive Facility at \url{http://archive.eso.org/cms.html}. The $YJ \ks$ tiles and PSF catalogue used in this work are publicly available and can be downloaded from the VSA at \url{http://vsa.roe.ac.uk/index.html} and ESO at \url{https://archive.eso.org/cms/eso-archive-news/new-data-release-of-the-vista-magellanic-cloud-survey-vmc.html}.
\bibliographystyle{mnras}
\bibliography{references}

\appendix

\bsp	
\label{lastpage}
\end{document}